\documentclass[10pt,a4paper,english]{article}\usepackage[]{graphicx}\usepackage[]{xcolor}
\makeatletter
\def\maxwidth{ %
  \ifdim\Gin@nat@width>\linewidth
    \linewidth
  \else
    \Gin@nat@width
  \fi
}
\makeatother

\definecolor{fgcolor}{rgb}{0.345, 0.345, 0.345}

\usepackage{framed}
\makeatletter
 {\par\unskip\endMakeFramed%
 \at@end@of@kframe}
\makeatother

\definecolor{shadecolor}{rgb}{.97, .97, .97}
\definecolor{messagecolor}{rgb}{0, 0, 0}
\definecolor{warningcolor}{rgb}{1, 0, 1}
\definecolor{errorcolor}{rgb}{1, 0, 0}
\newenvironment{knitrout}{}{} 

\usepackage{alltt}

\batchmode

\usepackage[hyphens]{url}
\usepackage{amsmath}
\usepackage{amsfonts}
\usepackage{amssymb}
\usepackage{amsthm}

\theoremstyle{plain}
\newtheorem{theorem}{Theorem}[section]

\newtheorem{corollary}[theorem]{Corollary}

\theoremstyle{definition}

\theoremstyle{remark}

\PassOptionsToPackage{hyphens}{url}\usepackage{hyperref}
\usepackage{hyperref}
\usepackage[square,numbers]{natbib}
\usepackage{orcidlink}
\usepackage{gitinfo2}

\usepackage[newitem,newenum,increaseonly]{paralist}


\usepackage[colorinlistoftodos]{todonotes}

\usepackage[notheorems]{SharpeR}

\providecommand{\altSNR}[0]{\xi}
\providecommand{\ptvsnr}[1][]{\vectUL{\altSNR}{}{#1}}
\providecommand{\stvsr}[1][]{\vectUL{\hat{\altSNR}}{}{#1}}
\providecommand{\ptsrUL}[2]{\mathUL{\altSNR}{#1}{#2}}
\providecommand{\ptsnr}[1][]{\ptsrUL{}{#1}}
\providecommand{\stsrUL}[2]{\mathUL{\hat{\altSNR}}{#1}{#2}}
\providecommand{\stsr}[1][]{\stsrUL{}{#1}}
\providecommand{\ssravg}[1][]{\mathUL{\bar{\prvSNR}}{}{#1}}
\providecommand{\psnrone}{\psnr[1]}
\providecommand{\psnrmax}{\psrUL{}{max}}

\usepackage{tikz}
\usetikzlibrary{arrows.meta, positioning, shapes.geometric}

    \usepackage{color} 
    \usepackage{enumerate} 
    \usepackage{amsmath} 
    \usepackage{amssymb} 
		\usepackage{fancyvrb} 
    \usepackage{hyperref}

    \definecolor{orange}{cmyk}{0,0.4,0.8,0.2}
    \definecolor{darkorange}{rgb}{.71,0.21,0.01}
    \definecolor{darkgreen}{rgb}{.12,.54,.11}
    \definecolor{myteal}{rgb}{.26, .44, .56}
    \definecolor{gray}{gray}{0.45}
    \definecolor{lightgray}{gray}{.95}
    \definecolor{mediumgray}{gray}{.8}
    \definecolor{inputbackground}{rgb}{.95, .95, .85}
    \definecolor{outputbackground}{rgb}{.95, .95, .95}
    \definecolor{traceback}{rgb}{1, .95, .95}
    \definecolor{red}{rgb}{.6,0,0}
    \definecolor{green}{rgb}{0,.65,0}
    \definecolor{brown}{rgb}{0.6,0.6,0}
    \definecolor{blue}{rgb}{0,.145,.698}
    \definecolor{purple}{rgb}{.698,.145,.698}
    \definecolor{cyan}{rgb}{0,.698,.698}
    \definecolor{lightgray}{gray}{0.5}
    
    \definecolor{darkgray}{gray}{0.25}
    \definecolor{lightred}{rgb}{1.0,0.39,0.28}
    \definecolor{lightgreen}{rgb}{0.48,0.99,0.0}
    \definecolor{lightblue}{rgb}{0.53,0.81,0.92}
    \definecolor{lightpurple}{rgb}{0.87,0.63,0.87}
    \definecolor{lightcyan}{rgb}{0.5,1.0,0.83}

    \definecolor{incolor}{rgb}{0.0, 0.0, 0.5}
    \definecolor{outcolor}{rgb}{0.545, 0.0, 0.0}

    \hypersetup{
      breaklinks=true,  
      colorlinks=true,
      urlcolor=blue,
      linkcolor=darkorange,
      citecolor=darkgreen,
      }
    
\providecommand{\RMAT}[1][{}]{\MtxUL{R}{}{#1}}

\providecommand{\AAA}[1][{}]{\MtxUL{A}{}{#1}}
\providecommand{\bbb}[1][{}]{\vectUL{b}{}{#1}}
\providecommand{\ccc}[1][{}]{\vectUL{c}{}{#1}}
\providecommand{\zzz}[1][{}]{\vectUL{z}{}{#1}}
\providecommand{\etav}[1][{}]{\vectUL{\eta}{}{#1}}
\providecommand{\yvec}[1][{}]{\vectUL{y}{}{#1}}
\providecommand{\Vfunc}[1]{\mathUL{\mathcal{V}}{#1}{}}
\providecommand{\Vmin}{\Vfunc{-}}
\providecommand{\Vmax}{\Vfunc{+}}
\providecommand{\tncdf}[5]{\funcit{F}{{#1};{#2},{#3},{#4},{#5}}}
\providecommand{\makerho}[3][{\vone}]{{#2}\wrapParens{\ogram{#1}} + {#3}\,\eye}
\providecommand{\eqscl}{c}
\providecommand{\pmone}[1][{}]{\vectUL{w}{}{#1}}
\providecommand{\posp}[1]{\mathUL{{#1}}{}{+}}
\providecommand{\pospsq}[1]{\mathUL{{#1}}{2}{+}}

\providecommand{\xxx}[1][{}]{\vectUL{x}{}{#1}}
\providecommand{\yyy}[1][{}]{\vectUL{y}{}{#1}}
\providecommand{\xxs}[1][{}]{\mathUL{x}{}{#1}}
\providecommand{\yys}[1][{}]{\mathUL{y}{}{#1}}

\IfFileExists{upquote.sty}{\usepackage{upquote}}{}
\begin{document}

\title{Conditional inference on the asset with maximum Sharpe ratio}
\author{\orcidlink{0000-0002-4197-6195} Steven E. Pav \thanks{\email{steven@gilgamath.com}.
The source code to build this document is available at
\href{http://www.github.com/shabbychef/maxsharpe}{\normalfont\texttt{www.github.com/shabbychef/maxsharpe}}.
The document you are reading was built from commit \texttt{877700c78d78ac58b3caf65d20f0f77926b2e904} of that repo.
}}

\maketitle

\begin{abstract}
We apply the procedure of Lee \etal \cite{lee2013exact} to the problem of performing
inference on the \txtSNR of the asset which displays maximum sample \txtSR over
a set of possibly correlated assets.
We find a multivariate analogue of the commonly used
approximate standard error of the \txtSR to use in this conditional estimation procedure.
We also consider several alternative procedures, including
the simple Bonferroni correction for multiple
hypothesis testing, together with a correction for positive common
correlation among assets under the null of equal \txtSNRs,
the chi-bar-square test against one-sided alternatives,
Follman's test, 
and Hansen's asymptotic adjustments.  \cite{nla.cat-vn3800977,10.2307/2291680,Hansen2003Asymptotic}

Testing indicates the conditional inference procedure achieves nominal type I rate,
and does not appear to suffer from non-normality of returns.
The conditional estimation test has low power under the alternative
where there is little spread in the \txtSNRs of the assets, 
and high power under the alternative where a single asset has high \txtSNR.
Unlike the alternative procedures, it appears to enjoy rejection probabilities
monotonic in the \txtSNR of the selected asset, 
and actually maintains near-nominal rejection rates under the conditional null.
\end{abstract}

\section{Introduction}

The problem of overfitting quantitative investment strategies
is certainly as old as the problem of selecting quantitative investment
strategies.
The choice of a course of action (\eg making an investment) based on 
historical observations leads to biased estimates of 
the value of the selected course of action when one uses the same historical
observations to estimate value.
That is, the estimates are ``biased by selection''.
This problem is not unique to quantitative finance, and goes
by many names: overfitting, p-hacking, data-mining bias, \etc
To be clear we are interested in the case where one has
observed \ssiz independent contemporaneous observations of returns
from \nstrat different ``assets'' (these can be trading strategy
backtest returns, or mutual fund returns, \etc), selects
one of those assets based on the historical performance, say
by selecting the asset with maximum \txtSR; then one wishes
to estimate or perform inference on the true `value' of the asset, for
example its \txtSNR, which we define as population analogue of the
\txtSR.

Aronson gives a good overview of the problem from the practicioner's
point of view, noting the relevant factors are the length of
history, the number of strategies tested, the correlation of their
historical performance, presence of outliers (or fat-tailedness of returns), and variation in
expected true effect size.  \cite[Chapter 6]{Aronson2007}
White's Reality Check was a pioneering development in the
area, giving a generally applicable method of estimating 
whether a selected model was superior to a benchmark model. \cite{White:2000}
White's work was extended and generalized by Romano and Wolf, Hansen, \emph{inter alia}. 
\cite{romano2005stepwise,Hansen:2005,Hsu2010471}
From a practical point of view the Reality Check and its variants do not scale
computationally to hundreds or thousands of assets, as they are based on
a (block) bootstrap. However, these methods can be adapted to very general
problems, can deal with correlation and autocorrelation of asset returns,
and are fairly robust to assumptions.

Recent work by L\'{o}pez de Prado and Bailey,
adapting standard techniques from Multiple Hypothesis Testing (MHT),
has gained attention 
in the field\footnote{Though application of MHT corrections
to the problem is not new: White's starting
assumption was apparently that such simple corrections
were inadequate. \cite{White:2000}}.  \cite{lopezdeprado2018false} 
They find the asymptotic expected value of
the maximum \txtSR of uncorrelated assets with zero \txtSNR.  
While use of simple techniques from MHT (Bonferroni correction, say)
can lead to reduced power, and is fragile with respect to assumptions,
they are alluring in their simplicity.
The Bonferroni correction is very simple to describe and implement,
and does not require one to store the historical returns of the
assets. It easily scales to millions of tested assets.

In this paper we exploit a result from Lee \etal on the problem
of \emph{conditional estimation}.  \cite{lee2013exact}
The Lee procedure was originally devised for analysis of the Lasso,
but is applicable in general to the case of selection from a 
normally distributed vector conditional to a linear constraint.
We simply give a multivariate normal approximation to the
vector of \txtSRs of \nstrat assets, then appeal to the Lee
\etal procedure.  

Our procedure is midway between the simple MHT
correction and the Reality Check tests, both computationally
and in robustness. Our procedure requires one to estimate
the correlation between returns, which would appear to require
\bigo{\ssiz \nstrat^2} runtimes. However, only the correlation of the
selected asset against all others is required, reducing the burden to \bigo{\ssiz\nstrat}.
Unlike the bootstrap tests, our procedure is easily adapted
to the case of producing confidence intervals on the 
\txtSNR, instead of only supporting hypothesis testing.


\section{Conditional Inference on the \txtSNR}

We consider the following problem: one has observed \ssiz \iid samples
of some $\nstrat$-vector \vreti, representing the returns of $\nstrat$
different ``assets,'' which could be stocks, trading strategies, \etc 
From the sample one computes the \txtSR of each asset, resulting in 
a $\nstrat$-vector, \svsr. One will then choose the asset with maximum
\txtSR. One then seeks to perform hypothesis tests or compute
confidence intervals on the \txtSNR of that asset. 
Here we define the \txtSR as the sample mean divided by sample
standard deviation, and the \txtSNR as the population analogue.
Throughout we use hats to denote sample quantities estimating
population parameters.

To simplify the exposition, we will suppose that, conditional 
on observing the vector \svsr, one rearranges the indices such
that the first asset has demonstrated the highest \txtSR.
This is to avoid the cumbersome notation of $\ssr[(1)]$,
and we instead can just write $\ssr[1]$. We note this maximum
condition can be written in the form $\AAA\svsr \le \bbb$ for
\bby{\nstrat-1}{\nstrat} matrix $\AAA$ defined by
$$
\AAA = \wrapBracks{\begin{array}{ccccc}
  -1 & 1 & 0 & \ldots & 0\\
  -1 & 0 & 1 & \ldots & 0\\
  \vdots & \vdots & \vdots & \ddots & \vdots\\
  -1 & 0 & 0 & \ldots & 1
\end{array}},
$$
and where $\bbb$ is the $\wrapParens{\nstrat-1}$-dimensional zero vector.
Also note that we are interested in performing inference on $\psnr[1]$,
which we can express as $\trAB{\etav}{\pvsnr}$ for $\etav=\basev[1]$.

Under these conditions, if only $\svsr$ were normally distributed, one
could use the following theorem due to Lee \etal:
\begin{theorem}[Lee \etal, Theorem 5.2 \cite{lee2013exact}]
\label{theorem:lee_etal}
Suppose $\yvec\sim\normlaw{\pvmu,\pvsig}$. 
  Define 
  $\ccc=\pvsig\etav / \qform{\pvsig}{\etav},$ and
  $\zzz=\yvec - \ccc\trAB{\etav}{\yvec}.$
  Let \pnorm[x] be the CDF of a standard normal, and 
  let \tncdf{x}{a}{b}{0}{1} be the CDF of a standard normal truncated
  to $\ccinterval{a}{b}$:
  $$
  \tncdf{x}{a}{b}{0}{1}\defeq\frac{\pnorm[x]-\pnorm[a]}{\pnorm[b] - \pnorm[a]}.
  $$
  Let \tncdf{x}{a}{b}{\pmu}{\psigsq} be the CDF of a general truncated normal,
  defined by
  $$
  \tncdf{x}{a}{b}{\pmu}{\psigsq} = \oneby{\psigma}
  \tncdf{\frac{x-\pmu}{\psigma}}{\frac{a-\pmu}{\psigma}}{\frac{b-\pmu}{\psigma}}{0}{1}.
  $$
  Then, conditional on $\AAA\yvec\le\bbb$, the random variable
  $$
  \tncdf{\trAB{\etav}{\yvec}}{\Vmin}{\Vmax}{\trAB{\etav}{\pvmu}}{\qform{\pvsig}{\etav}}
  $$
  is Uniform on $\ccinterval{0}{1}$, where \Vmin and \Vmax are given by
  \begin{align*}
    \Vmin &= \max_{j:\wrapParens{\AAA\ccc}_j < 0} \frac{\bbb[j] -
    \wrapParens{\AAA\zzz}_j}{\wrapParens{\AAA\ccc}_j},\\
    \Vmax &= \min_{j:\wrapParens{\AAA\ccc}_j > 0} \frac{\bbb[j] -
    \wrapParens{\AAA\zzz}_j}{\wrapParens{\AAA\ccc}_j}.
  \end{align*}
\end{theorem}

This theorem gives us a way to perform hypothesis tests, 
by comparing 
$\tncdf{\trAB{\etav}{\yvec}}{\Vmin}{\Vmax}{\trAB{\etav}{\pvmu}}{\qform{\pvsig}{\etav}}$
to some cutoff.
It also suggests a procedure for computing confidence intervals on 
$\trAB{\etav}{\pvmu}$, namely by univariate search for a value of
$\trAB{\etav}{\pvmu}$ such that 
$\tncdf{\trAB{\etav}{\yvec}}{\Vmin}{\Vmax}{\trAB{\etav}{\pvmu}}{\qform{\pvsig}{\etav}}$
is equal to some cutoff value.

In the following section we will show that the \svsr is \emph{approximately}
normally distributed.
In the following section we will examine whether the normal approximation 
is good enough to use the procedure of Lee \etal for testing the 
\txtSNR of the asset with maximum \txtSR.

\subsection{Normal approximation of the distribution of \txtSRs}

Here we derive the asymptotic distribution of \txtSR, following
Jobson and Korkie 
\emph{inter alia}. \cite{jobsonkorkie1981,lo2002,mertens2002comments,pav_ssc,pav_the_book}
Consider the case of \nstrat possibly correlated returns streams,
with each observation denoted by the $\nstrat$-vector \vreti.
Let \pvmu be the \nstrat-vector of population means, and let
\pvmom[2] be the \nstrat-vector of the uncentered second moments.
Let \pvsnr be the vector of \txtSNRs of the assets. Let \rfr be the
`risk free rate'. We have
$$
\pvsnr[i] = \frac{\pvmu[i] - \rfr}{\sqrt{\pvmom[2,i] - \pvmu[i]^2}}.
$$

Consider the $2\nstrat$ vector of \vreti, `stacked' with
\vreti squared elementwise, \vcat{\vreti}{\vreti^2}.
The expected value of this vector is \vcat{\pvmu}{\pvmom[2]};
let \pvvar be the variance of this vector, assuming it exists.

Given \ssiz observations of \vreti, consider the simple
sample estimate
$$
\vcat{\svmu}{\svmom[2]} \defeq
\frac{1}{\ssiz}\sum_{i}^{\ssiz} \vcat{\vreti}{\vreti^2}.
$$
Under the multivariate central limit theorem \cite{wasserman2004all}
\begin{equation}
\sqrt{\ssiz}\wrapParens{\vcat{\svmu}{\svmom[2]} - \vcat{\pvmu}{\pvmom[2]}}
\rightsquigarrow 
\normlaw{0,\pvvar}.
\label{eqn:mvclt}
\end{equation}

Let \svsr be the sample \txtSR computed from the estimates \svmu and
\svmom[2]: 
$\svsr[i] = \fracc{\wrapParens{\svmu[i]-\rfr}}{\sqrt{\svmom[2,i] - \svmu[i]^2}}.
$
By the multivariate delta method, 
\begin{equation}
\sqrt{\ssiz}\wrapParens{\svsr - \pvsnr} 
\rightsquigarrow 
\normlaw{0,\qoform{\pvvar}{\wrapParens{\dbyd{\pvsnr}{\vcat{\pvmu}{\pvmom[2]}}}}}.
\label{eqn:delmethod}
\end{equation}
Here the derivative takes the form of two \sbby{\nstrat}
diagonal matrices pasted together side by side:
\begin{equation}
\begin{split}
\dbyd{\pvsnr}{\vcat{\pvmu}{\pvmom[2]}} 
&= 
\onebytwo{\Mdiag{\frac{\pvmom[2] - \pvmu\rfr}{\wrapParens{\pvmom[2] - \pvmu^2}^{3/2}}}}{ 
\Mdiag{\frac{\rfr-\pvmu}{2 \wrapParens{\pvmom[2] - \pvmu^2}^{3/2}}}},\\
&=  
\onebytwo{\Mdiag{\frac{\pvsigma + \pvmu\pvsnr}{\pvsigma^2}}}
{\Mdiag{\frac{- \pvsnr}{2 \pvsigma^2}}}.
\end{split}
\label{eqn:sr_deriv}
\end{equation}
where $\Mdiag{\vect{z}}$ is the matrix with vector \vect{z} on its diagonal,
and where the vector operations above are all performed elementwise,
where we define the vector 
$\pvsigma \defeq \wrapParens{\pvmom[2] - \pvmu^2}^{1/2}$,
with powers taken elementwise.

In practice, the population values, \pvmu, \pvmom[2], \pvvar
are all unknown, and so the asymptotic variance has to be estimated,
using the sample. 
This is impractical for large $\nstrat$, so instead one may wish to impose
some distributional assumptions on \vreti.


Consider the case where \vreti is drawn from a normal distribution with mean
\pvmu and covariance \pvsig.
Then, using Isserlis' Theorem \cite{Isserlis1918,HaldaneMoments},
we have
\begin{equation}
\label{eqn:elliptical_variances}
\pvvar=\twobytwo{\pvsig}{2\pvsig\Mdiag{\pvmu}}{2\Mdiag{\pvmu}\pvsig}{%
	2 \pvsig \hadm \pvsig
	+ 4 \Mdiag{\pvmu}\pvsig\Mdiag{\pvmu}
},
\end{equation}
where $\hadm$ denotes \emph{Hadamard multiplication}.

Let \RMAT be the correlation matrix of the returns, defined as
\begin{equation}
\RMAT\defeq\Mdiag{\pvsigma^{-1}}\pvsig\Mdiag{\pvsigma^{-1}},
\end{equation}
where $\pvsigma$ is the (positive) square root of the diagonal of \pvsig.
Then using the \pvvar given in \eqnref{elliptical_variances},
\eqnref{mvclt} becomes
\begin{equation}
\svsr \approx 
\normlaw{\pvsnr,\oneby{\ssiz}\wrapParens{\RMAT +
	\frac{1}{2} \Mdiag{\pvsnr} \wrapParens{\RMAT\hadm\RMAT} \Mdiag{\pvsnr}}}.
\label{eqn:apx_srdist_gaussian}
\end{equation}
(See the appendix.)
Note how in the case of scalar Gaussian returns, this reduces to the well
known standard error estimate of 
$\sqrt{\oneby{\ssiz}\wrapParens{1 + \half[\psnrsq]}}$.  
\cite{lo2002,mertens2002comments,baoestimation,pav_the_book}
In practice the correlation matrix \RMAT and the vector of \txtSNRs, \pvsnr,
have to be estimated and plugged in.

We claim that for the case of 
\emph{elliptically distributed} \vreti, 
\eqnref{apx_srdist_gaussian} can be generalized to 
\begin{equation}
\svsr \approx 
\normlaw{\pvsnr,\oneby{\ssiz}\wrapParens{\Mtx{R} +
	\frac{\kurty-1}{4} \ogram{\pvsnr} + 
	\frac{\kurty}{2} \Mdiag{\pvsnr} \wrapParens{\Mtx{R}\hadm\Mtx{R}} \Mdiag{\pvsnr}}},
\label{eqn:apx_srdist_elliptical}
\end{equation}
where \kurty is the ``kurtosis factor'', equal to one third the kurtosis of the
marginals.  \cite{vignat2007extension} 
However, elliptically distributed returns have no skew, which makes them
less than ideal for modeling returns series. 
Once again note how this equation reduces to the form of the standard error
described by Mertens in the case of $\nstrat=1$. \cite{mertens2002comments}



\begin{corollary}[to \theoremref{lee_etal}]
  \label{corollary:sr_est}
  Let $\vreti\sim\normlaw{\pvmu,\pvsig}$, with $\pvsnr=\pvmu \hadd \pvsigma$, where
  $\pvsigma=\vdiag{\pvsig}$.
  Let \RMAT be the correlation matrix.
  Suppose you observe \ssiz independent observations of \vreti then construct
  the \txtSR, \svsr.
  Then, conditional on $\AAA\svsr\le\bbb$, the random variable
  $$
  u=\tncdf{\trAB{\etav}{\svsr}}{\Vmin}{\Vmax}{\trAB{\etav}{\pvsnr}}{\qform{\Mtx{Q}}{\etav}}
  $$
  is Uniform on $\ccinterval{0}{1}$, where 
  \Vmin, \Vmax and 
  $\tncdf{x}{a}{b}{\pmu}{\psigsq}$
  are as in the theorem,
  and 
  $$
  \Mtx{Q} = \oneby{\ssiz}\wrapParens{\RMAT +
  \frac{1}{2} \Mdiag{\pvsnr} \wrapParens{\RMAT\hadm\RMAT} \Mdiag{\pvsnr}},
  $$
  as given in \eqnref{apx_srdist_gaussian}.

\end{corollary}

Note that the relationship between
\trAB{\etav}{\svsr} and $\Vmin$ and $\Vmax$ is such that $u$ is unlikely to be strictly monotonic increasing
with \trAB{\etav}{\svsr}, \emph{ceterus paribus}. 
However, when $\trAB{\etav}{\svsr}\to\infty$, we expect $u\to 1$, and so to test the null hypothesis
\begin{equation*}
\Hyp[0] : \trAB{\etav}{\pvsnr}=c\quad\mbox{versus}\quad \Hyp[1] :  \trAB{\etav}{\pvsnr}>c,
\end{equation*}
one should reject at the $\typeI$ level when 
$$
\tncdf{\trAB{\etav}{\svsr}}{\Vmin}{\Vmax}{c}{\qform{\Mtx{Q}}{\etav}} \ge 1 - \typeI.
$$

As stated the procedure requires that one estimate \Mtx{Q}, which requires one to estimate \RMAT and \pvsnr.
However, computing the test statistic only requires access to ${\Mtx{Q}}{\etav}$. 
In the main inferential task considered here, that vector is the 
covariance of the asset with maximum \txtSR against all the rest.

Note that \corollaryref{sr_est} has uses beyond the stated problem of performing inference on the asset with the largest \txtSR. 
For example, suppose you observe the \txtSRs of \nstrat assets, then select the asset with the largest \emph{absolute} \txtSR,
choosing whether to hold it long or short depending on the sign of the \txtSR.
You wish to perform inference on your strategy.
In this case, again reorder the assets such that the first asset has the highest absolute \txtSR, but also
flip the signs of the asset returns as necessary such that all assets have positive \txtSR. 
Then proceed as in the usual case, but add to $\AAA$ and $\bbb$ the conditional restriction that all
elements of $\svsr$ are non-negative.

One wishes to also use the result for more general problems wherein one will hold a \emph{portfolio}
of assets, conditional on some observed properties of \svsr. For example:
\begin{itemize}
  \item
    Suppose you observe the \txtSRs of \nstrat assets, then select the top $m$ by \txtSR, then
    you choose to hold an some portfolio of those $m$ assets.
    In this case set $\AAA$ and $\bbb$ to reflect the ``$m$ choose $\nstrat - m$'' relevant
    inequalities to condition on.
  \item
    Suppose you observe the \txtSRs of \nstrat assets, then select all assets with \txtSR greater than
    some minimum value, \psnr[*]. Then you choose to hold some portfolio of all assets that
    pass the bar. In this case you need to modify $\AAA$ and $\bbb$ to condition on the passing
    assets having \txtSRs greater than \psnr[*] and the remaining assets having lower \txtSR.
\end{itemize}
In these cases, the test vector \etav should reflect the chosen portfolio,
but the \txtSNR of a portfolio is \emph{not} the portfolio-weighted sum (or average) of the
\txtSNRs of the constituent assets. Indeed the \txtSNR of dollar-weighted portfolio \pportw is
$\fracc{\trAB{\pvmu}{\pportw}}{\sqrt{\qform{\pvsig}{\pportw}}}.$
However, if \pportw is expressed in \emph{volatility units}, then the \txtSNR is
$\fracc{\trAB{\pvsnr}{\pportw}}{\sqrt{\qform{\RMAT}{\pportw}}}.$ Thus assuming you
can estimate volatility (and \RMAT) without error\footnote{Typically the error in 
a volatility estimate is less critical than error in the estimate of the mean.  \cite{chopra1993effect}}, 
then one could transform a dollar denominated portfolio into a volatility denominated portfolio.
From this one can perform inference using the test vector $\etav=\fracc{\pportw}{\sqrt{\qform{\RMAT}{\pportw}}}$.

One could also use the procedure to test the hypothesis that the asset with maximum 
\txtSR has higher \txtSNR than the \emph{average} \txtSNR of all assets considered. 
This is the null commonly tested by the Reality Check and its variants.
It may be of limited practical utility, however, since the selected asset may still
have inferior \txtSNR.


\section{Alternative Approaches}
\label{sec:alternative_approaches}

Before considering alternative approaches, it is worthwhile to step back and illustrate how they fundamentally differ from the conditional approach considered above.
First, in \figref{quant_flowchart_I} we illustrate the observable flow chart of a practicing quantitative strategist.
By some process the quant specifies a collection of strategies to test, then backtests them.
We assume that the strategy with the highest observed \txtSR is then selected as a candidate for trading.
Some kind of statistical test is performed on the whole collection of observed backtests.
Depending on the outcome of that test, the selected strategy is either traded or not\footnote{Presumably if strategy does not pass the test, the quant starts all over again, or pursues a different career.}.

\begin{figure}[ht]
\centering
\begin{tikzpicture}[
    node distance=0.5cm and 0.5cm,
    process/.style={rectangle, draw, thick, minimum width=3.5cm, minimum height=0.8cm, align=center},
    outcome/.style={rectangle, draw, thick, minimum width=3.0cm, minimum height=1.2cm, align=center},
    arrow/.style={-Stealth, thick}
]

\node[process] (specify) {SPECIFY STRATEGIES};
\node[process] (backtest) [below=of specify] {BACKTEST STRATEGIES};
\node[process] (identify) [below=of backtest] {IDENTIFY BEST\\BACKTESTED STRATEGY};
\node[process] (test) [below=of identify] {PERFORM\\STATISTICAL TEST};

\node[outcome] (fail) [below left=0.6cm and -1.0cm of test] {TEST FAILS\\TO REJECT NULL};
\node[outcome] (rej) [below right=0.6cm and -1.0cm of test] {TEST\\REJECTS NULL};

\node[outcome] (surrender) [below=of fail] {GIVE UP?};
\node[outcome] (trade) [below=of rej] {TRADE STRATEGY.};

\draw[arrow] (specify) -- (backtest);
\draw[arrow] (backtest) -- (identify);
\draw[arrow] (identify) -- (test);

\draw[arrow] (test) -- (fail);
\draw[arrow] (test) -- (rej);
\draw[arrow] (fail) -- (surrender);
\draw[arrow] (rej) -- (trade);

\end{tikzpicture}
  \caption{A quant strategist's statistical workflow is illustrated.}
  \label{fig:quant_flowchart_I}
\end{figure}

However, this flowchart illustrates only the outwardly observable parts of this workflow.
In \figref{quant_flowchart_II} we include some of the latent states of this process
and mark some of the unconditional probabilities of events.
We treat the outcome of the strategy specification as potentially random, with
some probability $\alpha_0$ of yielding only ``bad'' strategies (those with \txtSNR below the
threshold value), and probability $\alpha_1 = 1-\alpha_0$ of yielding some ``good'' strategies.
The quant then backtests all the strategies and selects the one with the best observed \txtSR.
There is some probability that this may be a good or bad strategy. 
The total unconditional probability that the selected strategy is bad is $\beta_0 + \beta_1$,
while the probability of a good strategy is $\beta_2 = 1 - \beta_0 - \beta_1$.
We then imagine the outcome of some statistical test, which may reject or fail to reject
the null hypothesis.
The probabilities of these events are illustrated in the figure and we denote their
unconditional probabilities as $\gamma_0$ through $\gamma_5$.

\begin{figure}[ht]
\centering
\begin{tikzpicture}[
    node distance=0.5cm and 0.25cm,
    process/.style={rectangle, draw, thick, minimum width=2.5cm, minimum height=0.8cm, align=center},
    outcome/.style={rectangle, draw, thick, minimum width=1.7cm, minimum height=1.0cm, align=center},
    arrow/.style={-Stealth, thick}
]

\node[process] (specify) {SPECIFY STRATEGIES};
\node[process] (all_bad) [below left=1.0cm and -0.7cm of specify] {ALL\\STRATEGIES\\ARE BAD};
\node[process] (some_ok) [below right=1.0cm and -0.7cm of specify] {SOME\\STRATEGIES\\ARE GOOD};
\node[process] (backtest_bad) [below=of all_bad] {BACKTEST STRATEGIES\\SELECT BEST};
\node[process] (backtest_ok) [below=of some_ok] {BACKTEST STRATEGIES\\SELECT BEST};

\node[process] (bad_bad) [below=of backtest_bad] {SELECTED\\STRATEGY\\IS BAD};
\node[process] (bad_ok) [below left=0.5cm and -2.1cm of backtest_ok] {SELECTED\\STRATEGY\\IS BAD};
\node[process] (good_ok) [below right=0.5cm and -1.9cm of backtest_ok] {SELECTED\\STRATEGY\\IS GOOD};

\node[outcome] (fail_bad_bad) [below left=0.7cm and -0.5cm of bad_bad] {TEST\\FAILS TO\\REJECT\\NULL};
\node[outcome] (rej_bad_bad) [below right=0.7cm and -1.95cm of bad_bad] {TEST\\REJECTS\\NULL};

\node[outcome] (fail_bad_ok) [below left=0.7cm and -0.6cm of bad_ok] {TEST\\FAILS TO\\REJECT\\NULL};
\node[outcome] (rej_bad_ok) [below right=0.7cm and -1.85cm of bad_ok] {TEST\\REJECTS\\NULL};
\node[outcome] (fail_good_ok) [below left=0.7cm and -1.4cm of good_ok] {TEST\\FAILS TO\\REJECT\\NULL};
\node[outcome] (rej_good_ok) [below right=0.7cm and -1.05cm of good_ok] {TEST\\REJECTS\\NULL};

\draw[arrow] (specify) -- node[right, xshift=0.2cm] {$\alpha_0$} (all_bad);
\draw[arrow] (specify) -- node[right, xshift=0.2cm] {$\alpha_1=1-\alpha_0$} (some_ok);
\draw[arrow] (all_bad) -- (backtest_bad);
\draw[arrow] (some_ok) -- (backtest_ok);
\draw[arrow] (backtest_bad) -- node[right, xshift=0.2cm] {$\beta_0=\alpha_0$} (bad_bad);
\draw[arrow] (backtest_ok) -- node[right, xshift=0.2cm] {$\beta_1$} (bad_ok);
\draw[arrow] (backtest_ok) -- node[right, xshift=0.2cm] {$\beta_2=1-\beta_0-\beta_1$} (good_ok);

\draw[arrow] (bad_bad) -- node[left] {$\gamma_0$} (fail_bad_bad);
\draw[arrow] (bad_bad) -- node[right] {$\gamma_1$} (rej_bad_bad);

\draw[arrow] (bad_ok) -- node[left] {$\gamma_2$} (fail_bad_ok);
\draw[arrow] (bad_ok) -- node[right] {$\gamma_3$} (rej_bad_ok);

\draw[arrow] (good_ok) -- node[left] {$\gamma_4$} (fail_good_ok);
\draw[arrow] (good_ok) -- node[right] {$\gamma_5=1-\gamma_0-\ldots-\gamma_4$} (rej_good_ok);

\end{tikzpicture}

  \caption{A quant strategist's statistical workflow with latent states is illustrated.
The unconditional probabilities in each row sum to $1.0$. The classical Bonferroni procedure
defines the rate of type I errors as $\gamma_1 / \beta_0$, in contrast to the conditional test.}
  \label{fig:quant_flowchart_II}
\end{figure}

The point of this figure is to illustrate the difference between the conditional procedure and the
procedures below, which test a different null hypothesis than the conditional procedure.
The conditional procedure approximately controls the rate of type I errors which can be written as
$$
\frac{\gamma_1 + \gamma_3}{\beta_0 + \beta_1}.
$$
The power of the conditional procedure is the ratio $\gamma_5 / \beta_2$.

The Bonferroni procedure, and indeed all the procedures introduced in this section, only provide
control over the type I rate $\gamma_1 / \beta_0$. 
The power of these procedures is 
$$
\frac{\gamma_3 + \gamma_5}{\beta_1 + \beta_2}.
$$
There is considerable tension between the power and type I rate of this null versus that of the conditional procedure.
Indeed in simulations we will find instances where $\gamma_3 / \beta_1 \approx 1.0$ for some tests,
although $\beta_1$ will be small for these cases.

Using this diagram we can also easily illustrate what \citeauthor{10.1111/1467-9868.00346} called the 
\emph{positive False Discovery Rate} (pFDR).  \cite{10.1111/1467-9868.00346}
The pFDR is the conditional probability of a false discovery conditional on the statistical test having rejected the null.
In our diagram it is computed as
$$
pFDR = \frac{\gamma_1 + \gamma_3}{\gamma_1 + \gamma_3 + \gamma_5}.
$$
We would argue that pFDR is more important to control than the type I rate of a test.
However, the pFDR depends strongly on the configuration of the population values, so
it is hard to quantify it in a way that is relevant in all situations.
Similarly it is hard to describe and test the null of the conditional procedure.
The Bonferroni null, however, is easy to describe and test: simply set the \txtSNR of every strategy
to the null value.
That is, the population \pvsnr takes only one value under the Bonferroni null,
but can have many different values under the conditional null.
This will become more clear in the section on simulations under the null and alternative.

\subsection{Bonferroni correction with simple correlation fix}

\label{subsec:fix_bonferroni}
The simple MHT approach to the problem is via a Bonferroni correction.  \cite{bretz2016multiple}
In its usual form, it assumes that the returns \vreti are independent
and normally distributed.  
In this case the marginals of \svsr are independent, and distributed
as rescaled non-central \tstat{} random variables.
So to test the null hypothesis
\begin{equation}
\label{hyp:eq_test}
\Hyp[0] : \forall_i \psnr[i]=\psnr[0]\quad\mbox{versus}\quad \Hyp[1] :  \exists_i \psnr[i] > \psnr[0],
\end{equation}
compute the \txtSRs, \svsr, then
reject at the $\typeI$ level when
$\sqrt{\ssiz} \max_i \svsr[i]$ exceeds $\nctqnt{1-\typeI/\nstrat}{\sqrt{\ssiz} \psnr[0]}{\ssiz-1}$,
the $1-\typeI/\nstrat$ quantile of the 
non-central \tstat-distribution with $\ssiz - 1$
degrees of freedom and non-centrality parameter $\sqrt{\ssiz} \psnr[0]$.

Note that the null tested here is \emph{not} the traditional one-sided null which is of
interest to the practicing quant, namely
\begin{equation}
\label{hyp:ineq_test}
\Hyp[0,s] : \forall_i \psnr[i] \le \psnr[0].
\end{equation}
The uncorrected Bonferroni test is valid for both: it rejects when $\max_i \ssr[i]$ is large,
so its rejection rate is nondecreasing in each \psnr[i].
The corrections for correlation we introduce below, however, are not valid for both:
for $\rho > 0$ they have negative weight on the average \txtSR, and so may reject
at well above the nominal rate when some $\psnr[i] < \psnr[0]$,
even though none exceeds \psnr[0].
We have confirmed this for the corrected Bonferroni test of \eqnref{fix_bonf_z}.
The chi-bar-square, Follman and Hansen tests built on the transformed \txtSRs of
\eqnref{xi_transform} share the same structure, and will likely not deliver the nominal type I rate under the null $\Hyp[0,s]$
of (\ref{hyp:ineq_test}).
A rejection by these tests is evidence against equality, not evidence that some $\psnr[i] > \psnr[0]$.
Another way of putting this is that the usual least favorable configuration argument will not apply to these tests.

This simple test is conservative, rejecting at below the nominal type I rate, in the presence of 
correlated assets. This can be demonstrated empirically, as we do in the sequel. 
One can also get a theoretical hint of why this holds by considering 
the normal approximation of \svsr given in \eqnref{apx_srdist_gaussian},
then appealing to Slepian's Lemma.
Slepian's Lemma establishes that for a normally distributed Gaussian vector
with fixed mean and variance, the maximum element is `stochastically decreasing' as
correlations increase.  \cite{slepian1962one}
Intuitively the number of true independent assets is decreasing as correlation
increases.

Let us consider a simple model for the correlation matrix
\begin{equation}
\label{eqn:simple_RMAT}
\RMAT=\makerho{\rho}{\wrapParens{1-\rho}},
\end{equation}
where $\abs{\rho} \le 1$.
Now simplify \eqnref{apx_srdist_gaussian} to 
\begin{equation}
\svsr \approx \normlaw{\pvsnr,\oneby{\ssiz}\RMAT},
  \label{eqn:apx_srdist_simple}
\end{equation}
which is reasonable in the case of the small \txtSNRs likely to be encountered in practice.
Then under the null hypothesis that $\pvsnr=\pvsnr[0]$, one observes
\begin{equation}
  \vect{z} = \sqrt{\ssiz}\ichol{\RMAT} \wrapParens{\svsr - \pvsnr[0]} \approx \normlaw{\vzero,\eye},
  \label{eqn:bonf_zform_simple}
\end{equation}
where $\ichol{\RMAT}$ is the inverse of the (symmetric) square root of $\RMAT$.

Under the assumed form for \RMAT given in \eqnref{simple_RMAT}, it is simple to confirm that
\begin{equation}
  \label{eqn:simple_RMAT_ichol}
  \ichol{\RMAT} = \makerho{\oneby{\nstrat}\wrapParens{\oneby{\sqrt{1-\rho+\nstrat\rho}} - \oneby{\sqrt{1-\rho}}}}{\wrapParens{1-\rho}^{-1/2}}.
\end{equation}
(This relation holds if we replace \ogram{\vone} in \RMAT with \ogram{\pmone} where
$\pmone$ is any vector whose elements are $\pm 1$. However in this case
we will lose the order-preserving property.)

Now it is simple to confirm that in this case the
transform induced by $\ichol{\RMAT}$ is ``order-preserving.''
That is, if $\vect{a} = \ichol{\RMAT}\vect{b}$ and $b_i \le b_j$ then $a_i \le a_j$.
As a consequence, if $i$ is the maximal element of $\wrapParens{\svsr - \pvsnr[0]}$,
then $i$ is the maximal element of $\vect{z}$.  Let us assume, again, that by
convention we have reordered the elements such that the first element of
\svsr is the maximum. Then to test the null hypothesis
$\forall_i \psnr[i]=\psnr[0]$, compute
\begin{equation}
  z_1 = \frac{\sqrt{\ssiz}\wrapParens{\ssr[1] - \psnr[0]}}{\sqrt{1-\rho}} + 
  \wrapParens{\oneby{\sqrt{1-\rho+\nstrat\rho}} - \oneby{\sqrt{1-\rho}}} \sqrt{\ssiz}\wrapParens{\frac{\tr{\vone}\svsr}{\nstrat} - \psnr[0]},
  \label{eqn:fix_bonf_z}
\end{equation}
and reject the null hypothesis when $z_1$ is bigger than 
$\qnorm{1-\typeI/\nstrat},$ the $1-\typeI/\nstrat$ quantile of the normal distribution.
In practice $\rho$ must be estimated. This could be done by computing the correlations
of the first asset against all others, then taking the average.

Note that the test statistic $z_1$ in \eqnref{fix_bonf_z} depends on elements of
\svsr other than \ssr[1]. 
Indeed it depends on the average value among the \ssr[i].
This may not be desireable, as it would reject as one of the \ssr[i] went to $-\infty$ for $i\ne 1$.
Moreover, the statistic $z_1$ does not seem to be entirely ``about'' \ssr[1], but
is computed from all elements of \svsr.
To rectify this, one is tempted to rotate the $\vect{z}$ from \eqnref{bonf_zform_simple}
to be maximally aligned with $\basev[1]$. 
This is an area of continued research.



Note that we did not have to make the simplifying assumption that led from 
\eqnref{apx_srdist_elliptical} to \eqnref{apx_srdist_simple} given the
form we assumed for \RMAT. That is, assuming $\RMAT=\makerho{a_2}{a_1}$,
then under the null hypothesis that $\pvsnr=\psnr[0]\vone$, 
\eqnref{apx_srdist_elliptical} becomes
\begin{equation}
\svsr \approx 
  \normlaw{\pvsnr,\oneby{\ssiz}\wrapParens{\makerho{a_2'}{a_1'}}},
\end{equation}
for some constants $a_1', a_2'$ which depend on $a_1, a_2, \kurty$ and $\psnr[0]$.
One could then proceed as above, constructing a $z_1$ statistic.

%
\paragraph{Bonferroni correction for arbitrary correlation structure}

The Bonferroni correction outlined above is strictly only applicable
to the rank-one correlation matrix, 
$\RMAT=\makerho{\rho}{\wrapParens{1-\rho}}$.
To apply the correction to any correlation matrix with positive entries,
Slepian's lemma allows us to appeal to a worst-case rank-one
correlation matrix.
\cite{slepian1962one,zeitouni2015gaussian}
Let $\xxx \sim \normlaw{\vzero,\RMAT}$ and
$\yyy \sim \normlaw{\vzero,\makerho{\rho}{\wrapParens{1-\rho}}}$,
for $\rho \ge 0$ where $\RMAT[i,j] \ge \rho$
for all $i\ne j$.
By Slepian's lemma, $\Pr{\max_i \xxs[i] \ge t} \le \Pr{\max_i \yys[i] \ge t}.$

Then assume the correlation of returns is \RMAT, 
where $\RMAT[i,j] \ge \rho$ for $i\ne j$ for some $\rho \ge 0$;
to test the null hypothesis
$\forall_i \psnr[i]=\psnr[0]$, compute $z_1$ as in \eqnref{fix_bonf_z},
plugging in $\rho$, 
and reject the null hypothesis when $z_1$ is bigger than 
$\qnorm{1-\typeI/\nstrat},$ the $1-\typeI/\nstrat$ quantile of the normal distribution.
This argument is heuristic: Slepian's lemma bounds the \emph{maximum} of a Gaussian vector,
but $z_1$ also depends on the average \txtSR, so we have not established that this
procedure has type I rate at most $\typeI$, even under the null of equality.


\subsection{Testing against one-sided alternatives}

Another obvious approach to the problem is to appeal to the
normal approximation of 
\eqnref{apx_srdist_gaussian} or \eqnref{apx_srdist_elliptical}, then
use well known techniques in testing of a multivariate normal
against a one-sided alternative.  \cite{nla.cat-vn3800977,vock2007}
That is, the usual multivariate procedure to test the null hypothesis 
$\Hyp[0] : \forall_i \psnr[i]=\psnr[0]$ under a normal approximation
would be via Hotelling's $T^2$ test.  \cite{anderson2003introduction,press2012applied}
However we are not interested in the case where some of
the $\psnr[i]$ are less than \psnr[0].

One-sided tests will not scale well to the case of large \nstrat, except perhaps
under simple models for correlation.
Consider testing the following null
\begin{equation*}
\Hyp[0] : \forall_i \psnr[i]=\psnr[0]\quad\mbox{versus}\quad \Hyp[1] :  
\forall_i \psnr[i]\ge\psnr[0]\,\,\mbox{and}\,\, \exists_j \psnr[j] > \psnr[0],
\end{equation*}
subject to the rank one correlation structure of \eqnref{simple_RMAT} with $\rho \ge 0$.
Again assume the approximation of \eqnref{apx_srdist_simple},
\begin{equation*}
\svsr \approx \normlaw{\pvsnr,\oneby{\ssiz}\RMAT}.
\end{equation*}
Letting $\stvsr = \ichol{\RMAT}\svsr$ and 
$\ptvsnr = \ichol{\RMAT}\pvsnr$, the normal approximation can be rewritten as
\begin{equation*}
  \sqrt{\ssiz}\stvsr \approx \normlaw{\sqrt{\ssiz}\ptvsnr,\eye}.
\end{equation*}
The inverse square root of \RMAT, given in \eqnref{simple_RMAT_ichol}, is order-preserving.
Moreover, 
\begin{equation*}
  \ichol{\RMAT} \vone = \eqscl\vone,
\end{equation*}
for $\eqscl=\wrapParens{1+\wrapParens{\nstrat-1}\rho}^{-\halff} > 0$.
From the order-preserving nature of \ichol{\RMAT}, the null and alternative hypotheses can be expressed
as 
\begin{equation*}
\Hyp[0] : \forall_i \ptsnr[i]=\eqscl \psnr[0]\quad\mbox{versus}\quad \Hyp[1] :  
\forall_i \ptsnr[i]\ge \eqscl \psnr[0]\,\,\mbox{and}\,\, \exists_j \ptsnr[j] > \eqscl \psnr[0].
\end{equation*}

We can then appeal to the simple chi-bar-square test.  \cite{nla.cat-vn3800977,wolak1987exact}
First transform the vector of \txtSRs to \stvsr via 
\begin{equation}
  \label{eqn:xi_transform}
\stvsr=\ichol{\RMAT}\svsr= \eqscl\ssravg\vone + \wrapParens{1-\rho}^{-\halff}\wrapParens{\svsr - \ssravg \vone},
\end{equation}
where $\ssravg$ is the average of the sample \txtSRs.
Then compute
\begin{equation}
  \label{eqn:chibstat}
  \bar{x}^2 = \ssiz \sum_i \pospsq{\wrapParens{\stsr[i] - \eqscl \psnr[0]}},
\end{equation}
where $\posp{y}$ is the positive part of $y$, \ie $\posp{y}=y$ if $y > 0$ and zero otherwise.
Then compute the CDF of the corresponding chi-bar-square distribution as
\begin{equation}
  \label{eqn:chibp}
  Q = \sum_{i=0}^{\nstrat} w_i {\chisqcdf{\bar{x}^2}{i}},
\end{equation}
where $\chisqcdf{x}{i}$ is the cumulative distribution of the $\chi^2$ distribution
with $i$ degrees of freedom, and $w_i$ are the chi-bar-square weights.
In this case they are defined as
$$
w_i = {\nstrat \choose i}{2^{-\nstrat}}.
$$
Reject the null hypothesis at the \typeI level if $1-Q \le \typeI$.

Note that, as with the Bonferroni correction, the test statistic $\bar{x}^2$ is
computed on all elements of \svsr, and thus the decision to reject the null
may not be ``about'' \ssr[1].
In testing we will see that the one-sided test is highly susceptible to distribution of the \pvsnr,
moreso than the Bonferroni correction.

We note that under this setup it is also easy to use Follman's test, which is a very simple procedure with
increased power against one-sided alternatives. \cite{10.2307/2291680}
Here we would compute
$$
g^2 = \ssiz \nstrat \eqscl^2 \wrapParens{\ssravg - \psnr[0]}^2 + \frac{\ssiz}{1-\rho}\sum_i \wrapParens{\ssr[i]-\ssravg}^2,
$$
and reject at the \typeI level if both $1-\chisqcdf{g^2}{\nstrat} \le 2 \typeI$ and $\ssravg > \psnr[0]$.

\subsection{Hansen's $\log \log$ Corrections}

One failing of many of the approaches considered above is the problem
of irrelevant alternatives.
That is, instead of testing under the null of equality, (\ref{hyp:eq_test}) above,
we should test the following
\begin{equation*}
\Hyp[0] : \forall_i \psnr[i] \le \psnr[0]\quad\mbox{versus}\quad \Hyp[1] :  \exists_i \psnr[i] > \psnr[0].
\end{equation*}
Testing such a composite null hypothesis is typically via a \emph{non-similar} test,
\ie one which has a type I rate no greater than the nominal rate for all \pvsnr in the null, 
and which achieves that nominal rate for some \pvsnr under the null. 
Such tests achieve the nominal rate under the \emph{least favorable configuration} (LFC),
which for the uncorrected Bonferroni test is the null of equality, (\ref{hyp:eq_test}),
but not for the corrections for correlation. \cite{nla.cat-vn3800977}

Hansen describes a procedure which avoids this problem. 
The idea is elegant, and ultimately simple to implement.  \cite{Hansen2003Asymptotic,Hansen:2005}
In the terms of the problem we consider, it amounts to assuming that the null takes the form
\begin{equation*}
  \Hyp[0] : \forall_i\,\psnr[i] \le \psnr[0]\mbox{ and }\abs{\psnr[i]-\ssr[i]} \le g_{\ssiz}\quad\mbox{versus}\quad \Hyp[1] :  \exists_i \psnr[i] > \psnr[0],
\end{equation*}
for some $g_{\ssiz}$. 
Note this seems odd since the sample \txtSR appears in the null hypothesis to be tested.
Hansen describes how such a test can be performed while maintaining a maximum
type I rate asymptotically, and achieving higher power.

Hansen applied this correction to the chi-bar-square statistic (\cf \eqnref{chibstat}),
and later to 
a Studentized version of White's Reality Check statistic, which is rather like
the corrected Bonferroni statistic computed in 
\eqnref{fix_bonf_z}.  \cite{Hansen2003Asymptotic,Hansen:2005}
Applying Hansen's correction to our problem is simple: compute \stvsr as in \eqnref{xi_transform}.
Let $\tilde{\nstrat}$ be the number of elements of \stvsr greater than $\eqscl\psnr[0] - \sqrt{\wrapParens{2\log\log\ssiz}/\ssiz}$,
where $\eqscl=\wrapParens{1+\wrapParens{\nstrat-1}\rho}^{-\halff}$.
If $\tilde{\nstrat}=0$ fail to reject.
Otherwise compute the chi-bar-square statistic $\bar{x}^2$ as in \eqnref{chibstat}
and reject if
\begin{equation*}
  \sum_{i=0}^{\tilde{\nstrat}} {\tilde{\nstrat} \choose i}{2^{-\tilde{\nstrat}}} {\chisqcdf{\bar{x}^2}{i}} \ge 1-\typeI.
\end{equation*}
We will refer to this as ``Hansen's chi-bar-square.'' 
It is the chi-bar-square test considered above, but with reduced degrees
of freedom \emph{which depend on the observed}.

The same correction is easily applied to the Bonferroni maximum test: again,
compute \stvsr and $\tilde{\nstrat}$. 
If $\tilde{\nstrat}=0$ fail to reject.
Otherwise reject at the $\typeI$ level if
\begin{equation*}
  \max_i \stvsr[i] - \eqscl\psnr[0] \ge \frac{\qnorm{1-\typeI/\tilde{\nstrat}}}{\sqrt{\ssiz}}.
\end{equation*}
We will refer to this as ``Hansen's SPA,'' although it does not
use the bootstrap procedure to estimate the standard error
as described by Hansen, it is similar in every other regard.  \cite{Hansen:2005}



\subsection{Subspace approximation}


Another potential approach to the problem which may be useful in
the case where returns are highly correlated, as one expects when
returns are from backtested quantitative trading strategies, is via 
a subspace approximation.
First we assume that the $\bby{\ssiz}{\nstrat}$ matrix of returns, \mreti
can be approximated by a $m$-dimensional subspace
$$
\mreti \approx \Mtx{Y}\Mtx{W},
$$
where $\Mtx{Y}$ is a $\bby{\ssiz}{m}$ matrix of `latent' returns,
and $\Mtx{W}$ is a $\bby{m}{\nstrat}$ `loading' matrix.

Now the column of \mreti with maximal \svsr has \txtSR that
is smaller than
$$
\ssropt \defeq \max_{\vect{\nu}} \frac{\trAB{\svmu}{\vect{\nu}}}{\sqrt{\qform{\svsig}{\vect{\nu}}}},
$$
where $\svmu$ is the $m$-vector of the (sample) means of columns of \Mtx{Y} and 
$\svsig$ is the sample covariance matrix.
This maximum takes value
$$
\ssropt = \sqrt{\qiform{\svsig}{\svmu}},
$$
which is, up to scaling, Hotelling's $T^2$ statistic.

Under the null hypothesis that the rows of \Mtx{Y} are independent draws from a Gaussian
random variable with zero mean, then
$$
\frac{\wrapParens{\ssiz-m}\ssrsqopt}{m \wrapParens{\ssiz-1}}
$$
follows an $F$ distribution with $m$ and $\ssiz-m$ degrees of 
freedom.  
Under the alternative it follows a non-central $F$ distribution.
\cite{anderson2003introduction,press2012applied}
Via this upper bound $\ssr[1] \le \ssropt$, one can then perform
tests on the null hypothesis $\forall_i \psnr[i]=0$.

However, this approach requires that one estimate $m$, the
dimensionality of the latent subspace. Moreover, the subspace
approximation may not be very good. It would seem that to
get near equality of \ssr[1] and \ssropt, the columns
of \mreti would have to contain both positive and negative
exposure to the columns of \Mtx{Y}. This in turn should result in 
mixed correlation of asset returns, which we may not observe
in practice. 
Finally, empirical testing indicates this approach requires further 
development.  \cite{pav_maxsharpe_two}


\subsection{Comparing Rejection Regions}

\begin{knitrout}\small
\definecolor{shadecolor}{rgb}{0.969, 0.969, 0.969}\color{fgcolor}\begin{figure}[h]
\includegraphics[width=0.975\textwidth,height=0.691\textwidth]{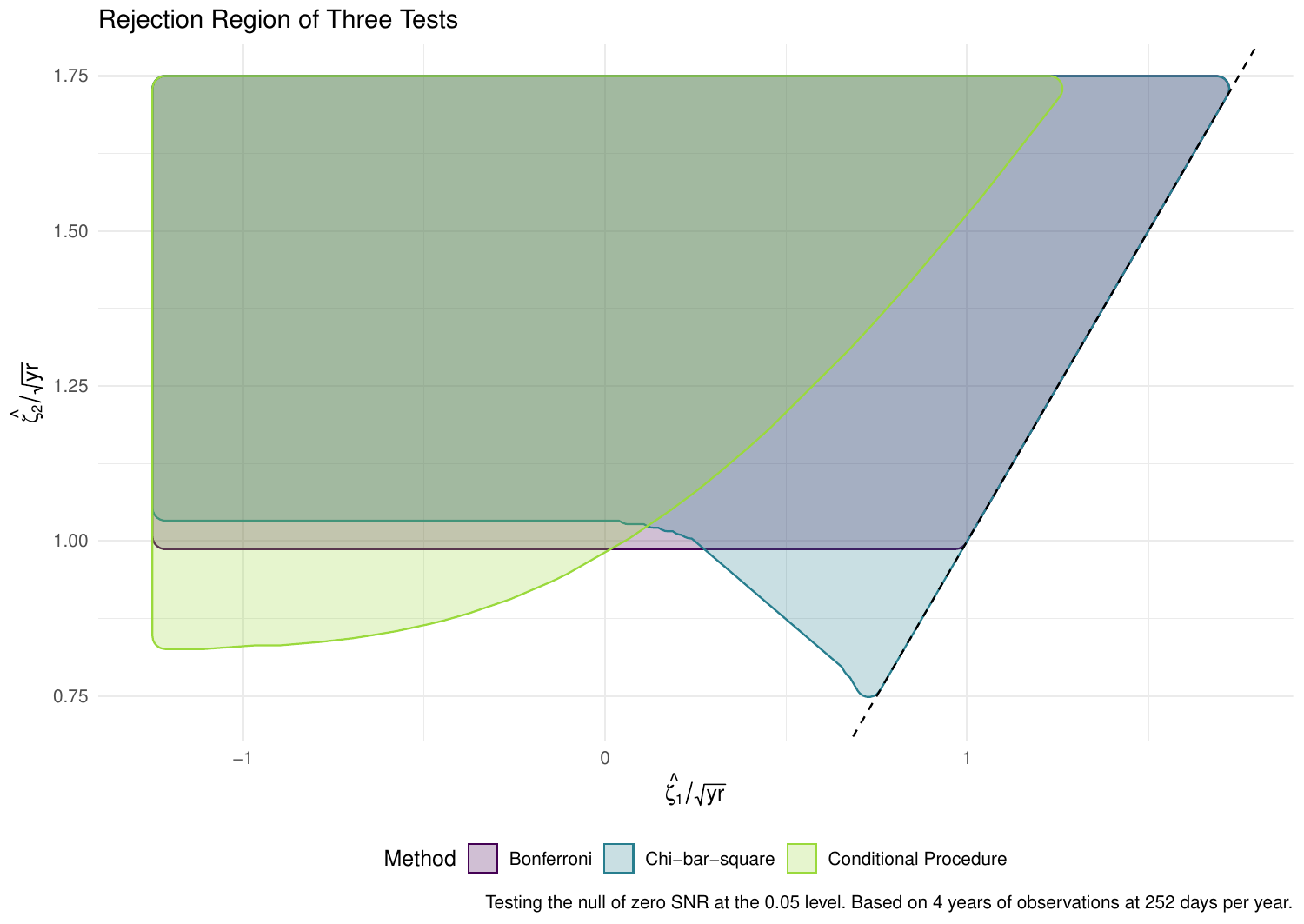} \caption[The 0.05 rejection regions for the vanilla Bonferroni adjusted test, the chi-bar-square test, and the conditional procedure are shown for the case of $\nstrat=2$ assets]{The 0.05 rejection regions for the vanilla Bonferroni adjusted test, the chi-bar-square test, and the conditional procedure are shown for the case of $\nstrat=2$ assets. Rejection regions are plotted for testing the null $\psnr[0] = 0$ based on 4 years of data at 252 days per year. The plot is conditional on $\ssr[2] \ge \ssr[1]$. }\label{fig:rejection_contours_plot}
\end{figure}

\end{knitrout}

Later in simulation we will see different performance from the conditional procedure and the procedures based on multiple hypothesis testing.
To give some sense of contrast between the tests we can look at the rejection regions for the simplest case, which is of two assets.
In \figref{rejection_contours_plot} we plot the 0.05 rejection regions for the Bonferroni correction, the chi-bar-square test, and the conditional test.
Each test is against the null $\psnr[0]=0$ based on $\ssiz=1,008$ days of data.

In that plot we see that the chi-bar-square test has slightly bigger rejection region than the Bonferroni procedure,
though it does not mean we are more likely to observe \svsr in that region under the alternative, as it depends
on the form of the alternative. 
Under their common null hypothesis the rejection regions should be hit only 5\% of the time.

We see that the conditional procedure does not reject when $\ssr[2] \approx \ssr[1]$.
When the two are close, the uniform variate that comes out of the polyhedral lemma should be close to zero and one would fail to reject.
The tradeoff is that conditional procedure appears to have a somewhat larger rejection region when $\ssr[1] < 0$.
We will observe later that the conditional procedure does enjoy higher power when there are very few very good strategies among those tested,
which is consistent with this plot.

\clearpage

\section{Empirical Results}

\subsection{Simulations under the null}


\subsubsection{Gaussian returns, infeasible estimator}

First we seek to establish if, and under what circumstances, 
the normal approximation of \eqnref{apx_srdist_gaussian} is sufficiently accurate to 
give nominal coverage under the conditional estimation procedure.
First we test a single case of $\nstrat=100$ using a correlation
matrix that is $\rho=0.7$ on the off-diagonals:
$\RMAT=\makerho{0.7}{0.3}$.
We generate Gaussian returns over $1260$ days, approximately
$5$ years worth for equity returns.
We let $\pvsnr$ range uniformly from $-0.1\dayto{-\halff}$ to
$0.1\dayto{-\halff}$. We compute the \txtSRs of
each asset's simulated returns, find the asset with maximum \txtSR,
then compute a p-value using \theoremref{lee_etal}. 
Since we wish to assess the accuracy of the normal approximation,
we use the actual population value of \RMAT, and the \pvsnr to compute
the covariance via \eqnref{apx_srdist_gaussian}.
This is not, of course, how the test would be applied in practice since
\RMAT and \pvsnr have to be estimated.
Moreover, we with to check coverage of the procedure under the null, so 
we use the actual $\trAB{\etav}{\pvsnr}$ in our test.

We repeat this experiment \ensuremath{10^{5}} times and collect the resultant putative p-values.
We would like to Q-Q or P-P plot these p-values, 
as evidence that they are near uniform, 
but the large sample size presents some challenges.
Instead, we choose some selected small $q$ (like 0.05 or 0.01), and
compute the proportion of our p-values $\le q$. We then subtract $q$.
This value, call it $\Delta$, should be near zero. We plot $\Delta$ against
$q$, with errorbars around the $x$ axis reflecting the area where we would
expect the points to fall roughly 95\% of the time. 
Those errorbars are computed via the Binomial law, but do not always have width
of exactly 95\% because of the finite sample size.
However, the plot suggests that the p-values are indeed uniformly distributed,
and that the procedure has near nominal type I rate when selecting a cutoff
near the selected $q$.

\begin{knitrout}\small
\definecolor{shadecolor}{rgb}{0.969, 0.969, 0.969}\color{fgcolor}\begin{figure}[h]
\includegraphics[width=0.975\textwidth,height=0.691\textwidth]{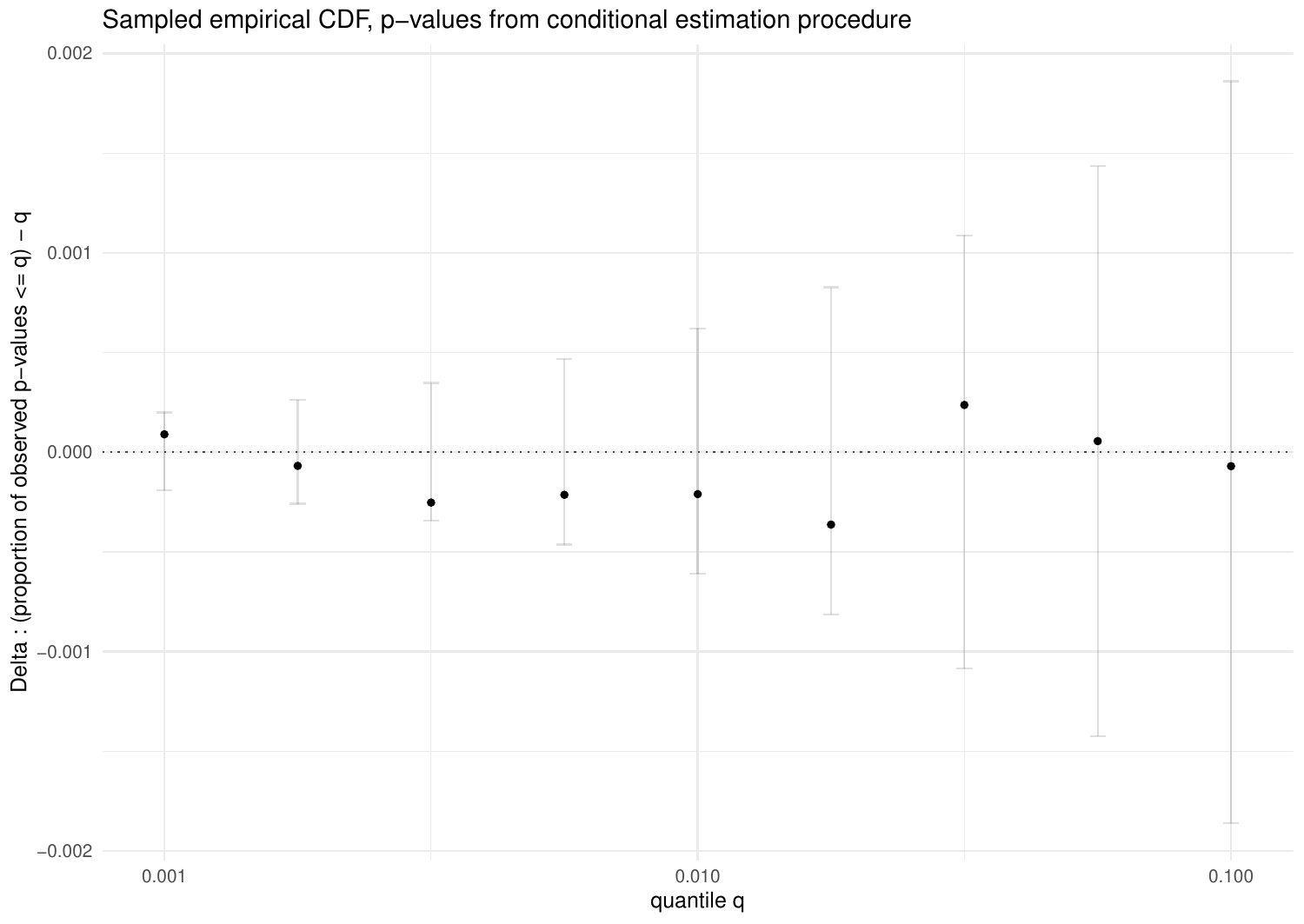} \caption[The computed p-values from the conditional estimation procedure over 1e+05 simulations are compared to a uniform law]{The computed p-values from the conditional estimation procedure over 1e+05 simulations are compared to a uniform law. Given the large sample size, a Q-Q plot is visually hard to interpret. Instead, for selected $q$, we compute the proportion of our putative p-values less than $q$. We then compute $\Delta = \wrapParens{\mbox{Prop }\le q} - q$ against $q$. Using the binomial law, we plot approximate error bars around the $x$ axis that indicate where the points should fall with approximately $95\%$ confidence. Simulations use the exact \RMAT and \pvsnr to compute the covariance matrix. The plot supports uniformity of the putative p-values. }\label{fig:motherload_sims_hidden_plotz}
\end{figure}

\end{knitrout}

\clearpage

\subsubsection{Gaussian returns, feasible estimator}

While these experiments suggest the normal approximation leads to nearly uniform p-values
under the null, they use the (unknown) population values of \RMAT and \pvsnr to compute
the variance-covariance of \svsr.
So we repeat the experiments, but plug in the usual sample estimate of covariance and the
vector of \txtSRs into \eqnref{apx_srdist_gaussian} to estimate the covariance
matrix of \svsr. 
Other than this change, we repeat the previous experiment, 
performing $\ensuremath{10^{5}}$ simulations, 
setting $\nstrat=100$,
$\RMAT=\makerho{0.7}{0.3}$.
$\ssiz=1260\dayto{}$, \etc
In \figref{causal_sims_log_plotz} we present a sampled CDF plot of the log p values, as above.
Again the simulations are consistent with the procedure having nominal coverage.
This is not surprising, because, as noted above, the statistical test
only requires us to estimate the standard error of the \txtSR of the
asset with maximum \txtSR, and so does not greatly rely on the $\nstrat^2$
elements of the estimate of \RMAT.

\begin{knitrout}\small
\definecolor{shadecolor}{rgb}{0.969, 0.969, 0.969}\color{fgcolor}\begin{figure}[h]
\includegraphics[width=0.975\textwidth,height=0.691\textwidth]{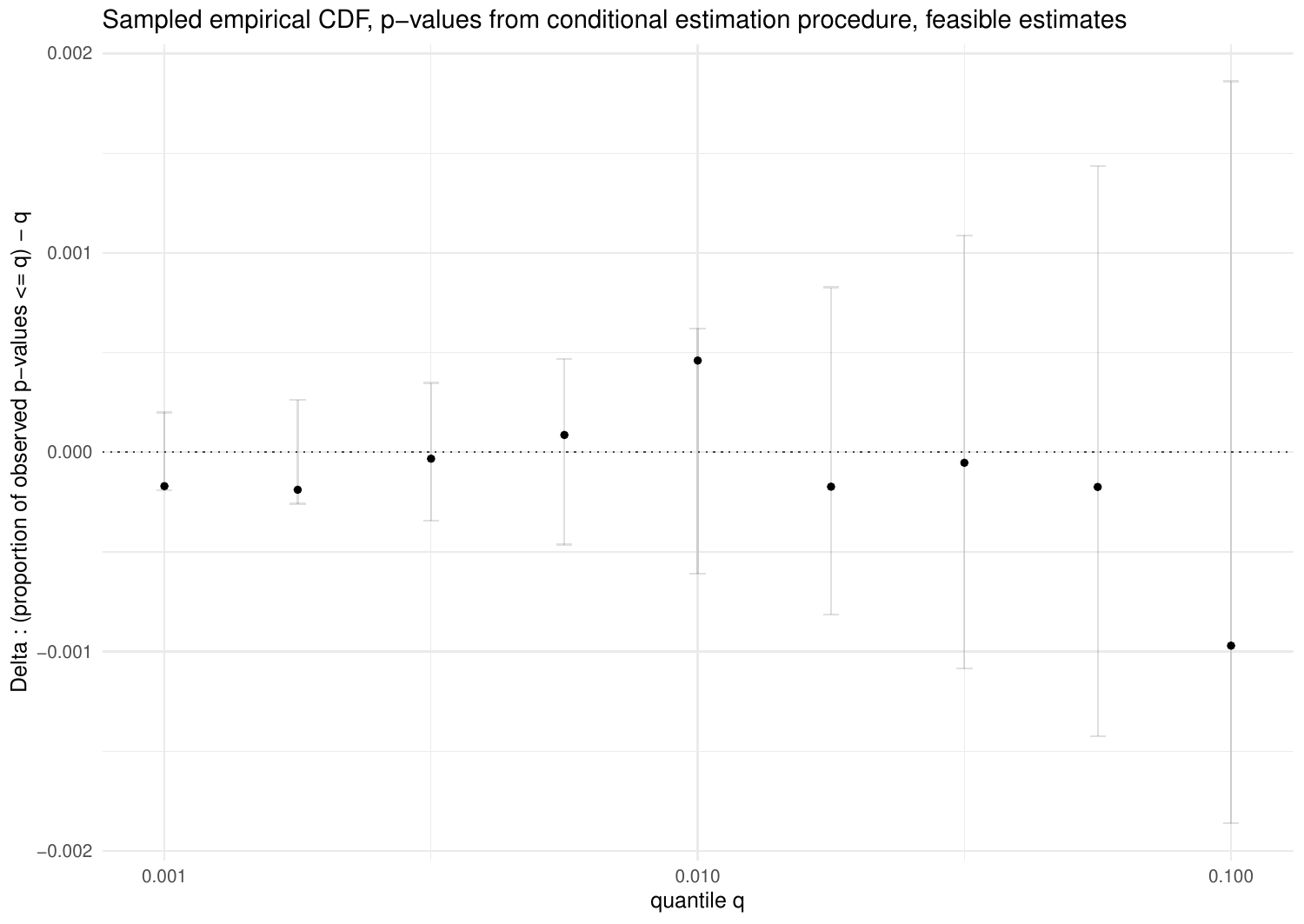} \caption[The computed p-values from the conditional estimation procedure using sample estimates of the covariance matrix, \RMAT, and the vector of \txtSNRs, \pvsnr are used in this sampled CDF plot]{The computed p-values from the conditional estimation procedure using sample estimates of the covariance matrix, \RMAT, and the vector of \txtSNRs, \pvsnr are used in this sampled CDF plot. P values are computed over 100,000 simulations. For selected $q$, we compute the proportion of our putative p-values less than $q$. We then compute $\Delta = \wrapParens{\mbox{Prop }\le q} - q$ against $q$. Using the binomial law, we plot approximate error bars around the $x$ axis that indicate where the points should fall with approximately $95\%$ confidence. Simulations use the exact \RMAT and \pvsnr to compute the covariance matrix. The plot supports uniformity of the putative p-values. }\label{fig:causal_sims_log_plotz}
\end{figure}

\end{knitrout}

\clearpage

\subsubsection{Gaussian returns, feasible estimator, sensitivity}


This kind of ``proof by eyeball'' is somewhat unsatisfying, and does not scale up to the task
of finding where the approximation is accurate.
To measure the uniformity of our p-values, we generate some via simulations as described above,
then compute the Kolmogorov-Smirnov statistic against a uniform distribution. 
\cite{Marsaglia:Tsang:Wang:2003:JSSOBK:v08i18}
You can think of the K-S statistic as the maximum absolute deviance of a point away
from the $y=x$ line in a P-P plot like 
\figref{motherload_sims_hidden_plotz}.

So we repeat the previous experiments, using a feasible estimator of the covariance matrix of \svsr. 
Again we draw returns from a Gaussian distribution.
We let $\ssiz$ vary from $63$ to $500$;
we let $\nstrat$ vary from $50$ to $200$;
we let $\rho$ vary from $0$ to $0.8$
where we take 
$\RMAT=\makerho{\rho}{\wrapParens{1-\rho}}$;
we take $\pvsnr$ to be a uniform sequence from $0$ to 
$0.1\dayto{-\halff}$.
For each setting of the parameters in the Cartesian product we perform
$\ensuremath{10^{5}}$ simulations, computing p values from the feasible estimator.

In \figref{many_sims_plotz} we plot those K-S statistics against $\ssiz$,
with different facets for $\rho$.
All else equal, we expect the approximation to be worse, and thus the K-S statistics
to be higher, for smaller \ssiz and larger \nstrat.
This pattern is somewhat visible in the plots, 
although large $\rho$ seems to reduce the number of `pseudo-assets' in that
relationship, and the number of observations, \ssiz, appears to have greater impact than the 
number of assets \nstrat.
However, with the given limited evidence,
we cannot claim to have definitively established where our procedure breaks down,
but warn users that the $\nstrat \gg \ssiz$ cases are likely to be problematic
in the sense that nominal type I rates may not be maintained.

\begin{knitrout}\small
\definecolor{shadecolor}{rgb}{0.969, 0.969, 0.969}\color{fgcolor}\begin{figure}[h]
\includegraphics[width=0.975\textwidth,height=0.691\textwidth]{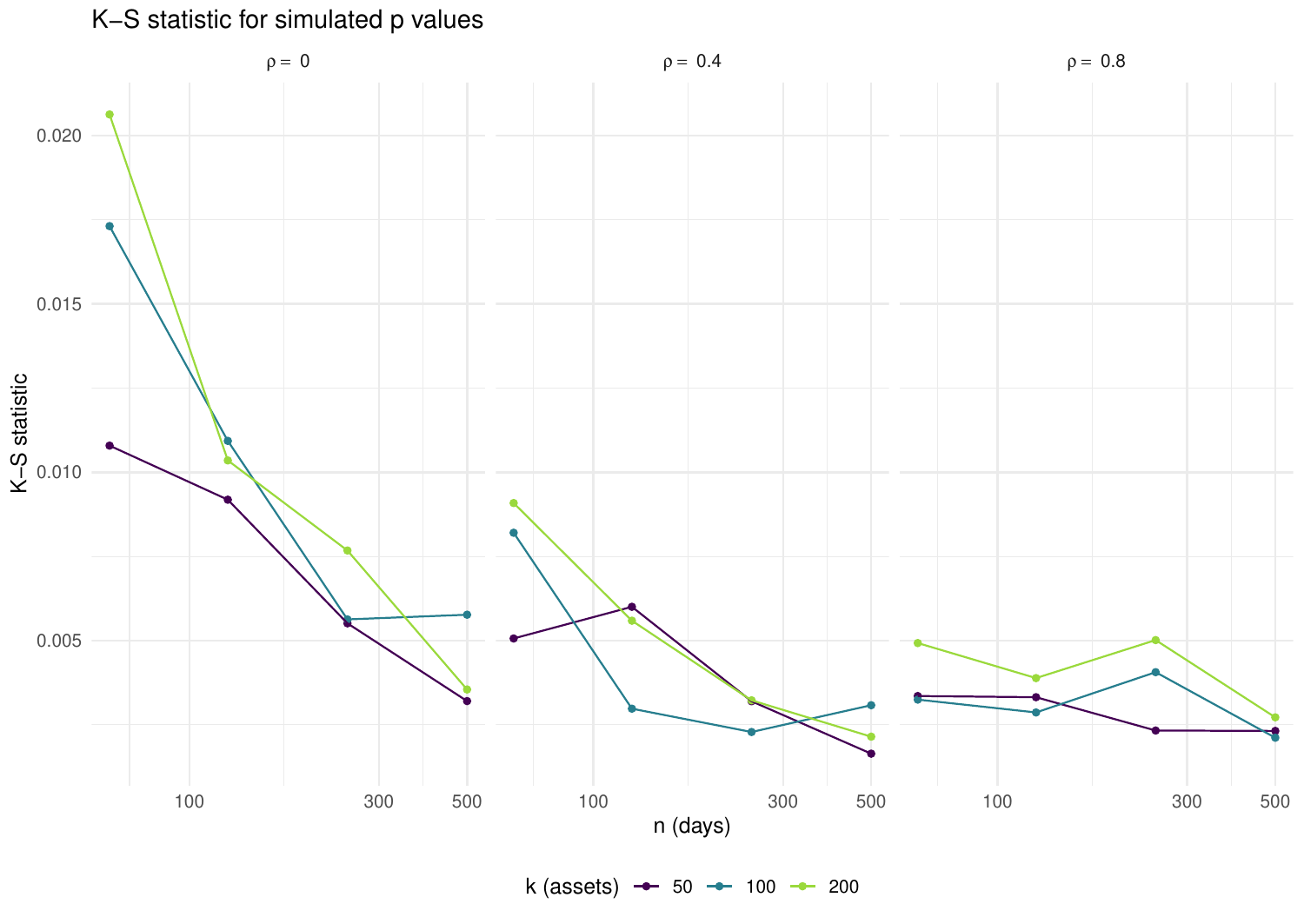} \caption[Kolmogorov-Smirnov statistics summarizing uniformity of the test statistic $u$ are plotted versus $\ssiz$ with facets for $\rho$]{Kolmogorov-Smirnov statistics summarizing uniformity of the test statistic $u$ are plotted versus $\ssiz$ with facets for $\rho$.  Broadly we see that the test statistic is less uniform in the regime where $\ssiz$ is small, but this is muted by a large positive correlation $\rho$.}\label{fig:many_sims_plotz}
\end{figure}

\end{knitrout}

\subsubsection{$\tstat{}$ returns, feasible estimator}

The simulations above were carried out assuming Gaussian returns, and using
\eqnref{apx_srdist_gaussian} to compute the covariance matrix of \svsr.
Gaussian returns are not a good model for real asset returns,
so we repeat those simulations with returns drawn from a multivariate $\tstat{}$-distribution
with $5$ degrees of freedom.  \cite{Lin1972339,kotz2004multivariate}
Again we perform $\ensuremath{10^{4}}$ simulations with
$\nstrat=100$,
$\RMAT=\makerho{0.7}{0.3}$,
$\ssiz=1260\dayto{}$, \etc
We perform inference twice, once using 
\eqnref{apx_srdist_gaussian},
and once using \eqnref{apx_srdist_elliptical} where we have estimated
the kurtosis factor, \kurty, by taking the median of the sample marginal 
kurtosises of the assets. 
In \figref{causal_tfive_sims_plotz} we present the subsampled empirical CDF plots
on the log-transformed p-values under the two methods of estimating
the covariance of \svsr.
There is little difference in the performance of the two sets of simulations,
though without the correction, the procedure is slightly anti-conservative, while with the correction
it is slightly conservative.
We remain cautiously optimistic that for large \ssiz, one need not correct
\eqnref{apx_srdist_gaussian} to account for non-normal returns.

\begin{knitrout}\small
\definecolor{shadecolor}{rgb}{0.969, 0.969, 0.969}\color{fgcolor}\begin{figure}[h]
\includegraphics[width=0.975\textwidth,height=0.691\textwidth]{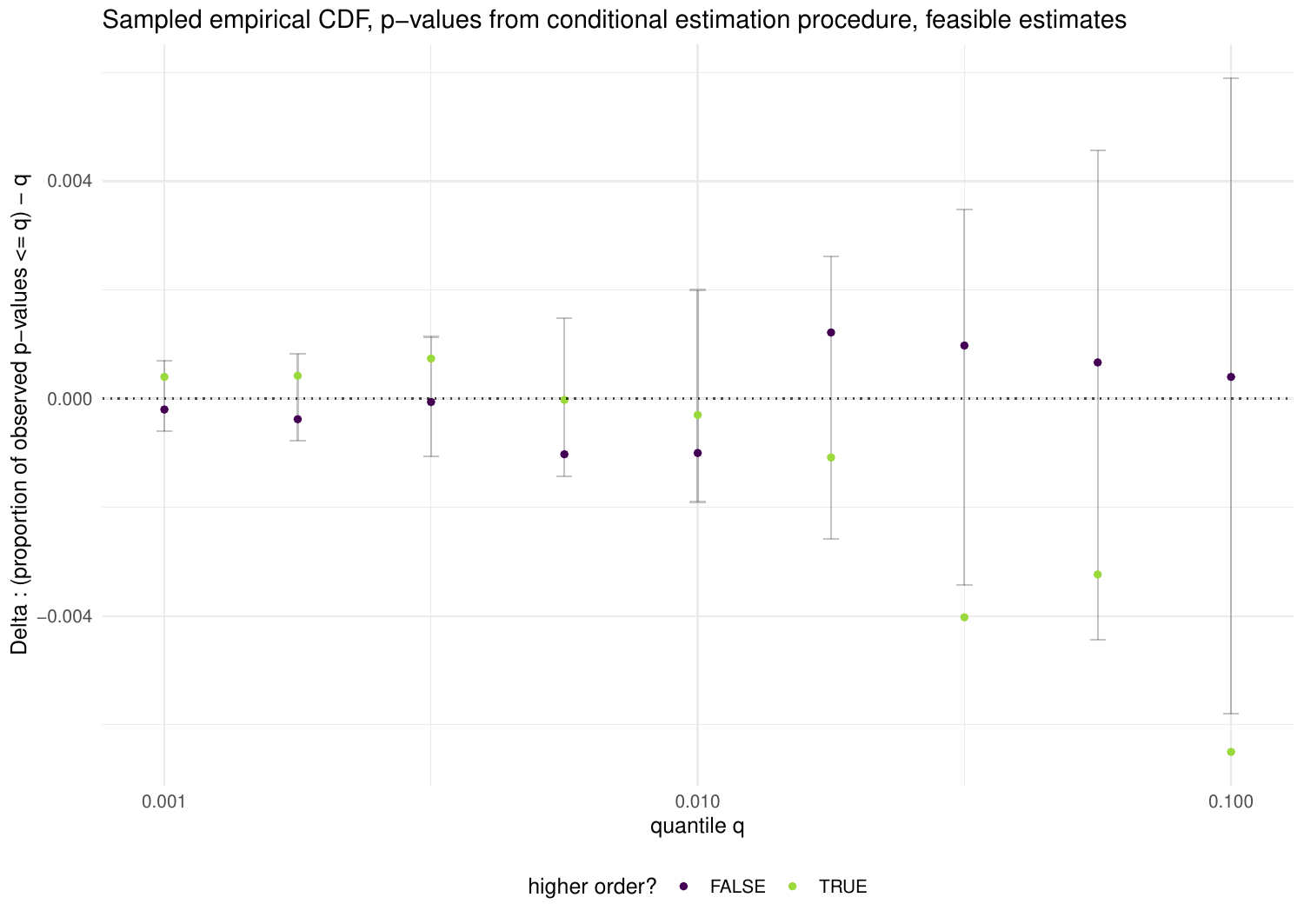} \caption[The computed p-values from the conditional estimation procedure over 10,000 simulations are plotted using our sampled CDF procedure]{The computed p-values from the conditional estimation procedure over 10,000 simulations are plotted using our sampled CDF procedure. Returns are drawn from a $\tstat{}\wrapParens{5}$ distribution.  Sample estimates are used to construct the variance-covariance matrix of \svsr.  The experiments are performed twice: first assuming that returns are Gaussian; then assuming returns are elliptical with unknown kurtosis factor that we estimate from the sample.  Both procedures give near-uniform p values. Without correcting for higher order moments, the procedure has higher than nominal type I rate; with the correction it has slightly smaller than the nominal rate. }\label{fig:causal_tfive_sims_plotz}
\end{figure}

\end{knitrout}

\clearpage

\subsection{Simulations under the alternative}
We wish to test the power of the method under the alternative hypothesis.
However, it is hard to state exactly what constitutes \emph{the} alternative.
One interpretation is that we condition on $\psnr[1] > 0$, where again
the indexing is such that $\ssr[1]$ was the maximum over $\nstrat$ assets;
then we estimate the probability of (correctly) rejecting $\psnr[1] = 0$
versus $\psnr[1]$. However, we suspect that the power, as described in 
this way, would depend on the distribution of
values of \pvsnr, as well as the correlation of returns among the tested strategies.

We will consider following alternatives: one where all \nstrat elements of \psnr are
equal (``all-equal''), and three others where $m$ of \nstrat elements of \psnr
are equal to some positive value, and the remaining $\nstrat - m$ are negative
that value. 
For $m=1$, we call this the ``one-good'' alternative; 
for $m=2$, the ``two-good'' alternative; 
for $m=\nstrat/2$, the ``half-good'' alternative. 
Later we also consider the cases where the elements of \psnr are
uniformly distributed across some finite set of values (``uniform'') or distributed like
the density of a scaled and shifted binomial distribution (``bell-shaped'').

We compare the power of the conditional estimation procedure to that of a
simple MHT correction, and the one-sided test.
In our experiments, we 
draw returns from a Gaussian distribution with diagonal covariance.
Under this assumption, one can use the distribution of the \tstat{} statistic
to perform inference on the \txtSNR. \cite{pav_ssc,pav_the_book}
We then use the Bonferroni correction to account for the multiple tests performed.
We also perform the one-sided test based on the chi-bar-square statistic,
Follman's test, and the chi-bar and MHT tests with Hansen's $\log\log$ adjustment.

Note that in the all-equal case, since every asset has the same \txtSNR,
whichever we select will have the same \txtSNR, and the Bonferroni-corrected test should have the
same power as the \tstat{}-test for a single asset. 
The conditional estimation procedure, however, may suffer in this case as
we may condition on a \ssr[1] that is very close to being non-optimal,
resulting in a small test statistic for which we do not reject.
On the other hand, for the one-good case, as the $\nstrat - 1$ assets 
may have considerably negative \txtSNR, they are unlikely to exhibit the
largest \txtSR, and so the MHT is merely testing a single asset, but
at the $\typeI / \nstrat$ level instead of the $\typeI$ level, resulting
in lower power. 
The chi-bar-square test and Follman's test are also unlikely to reject.
The conditional estimation procedure, however, should
not suffer under this alternative.
We also expect the $\log\log$ adjustment to have little effect
when the \txtSRs are all nearly equal, and more effect
when they are different.

Our suspicions are born out by the simulations.
We perform simulations under all-equal, half-good, two-good and one-good configurations, 
letting the `good' \txtSNR vary from 
$0$ to $0.212\dayto{-\halff}$, which corresponds
to an `annualized' \txtSNR of around
$3.4\yrto{-\halff}$.
We draw Gaussian returns with diagonal covariance for
$100$ assets, with $\ssiz=1008$.
For each setting we perform $\ensuremath{10^{4}}$ simulations then 
compute the empirical rejection rate of the test at the $0.05$ level,
conditional on the \txtSNR of the \emph{selected} asset, which is
to say the one with the largest \txtSR. Note that in some simulations
the largest \txtSR is observed for an asset with a negative \txtSNR.
We hope our tests to have lower power when this occurs.

In \figref{power_plotz_one_a}, we plot the power of 
Follman's test, 
Hansen's chi-bar-square and MHT (``SPA'') tests,
and the conditional estimation procedure versus the \txtSNR
of the selected asset. 
We present facet columns for the configurations of \pvsnr,
\viz all-equal, half-good, two-good and one-good.
A horizontal line at $0.05$ gives the nominal rate under the null,
which occurs as $x=0$ in these plots.

As expected from the above explanation, the
chi-bar-square (with Hansen's correction) has the highest power for the all-equal alternative,
followed by Follman's test, 
then the MHT, 
then the conditional estimation test.
These relationships are nearly exactly reversed for the one-good case:
Follman's test has zero power against the one-good alternative, 
while the conditional procedure has the highest power
Each test considered here shows similar performance in the half-good as in the all-equal alternative,
with the exception of Follman's test, which achieves a maximum power of $1/2$ in the
half-good case, as is to be expected since this is the probability that
$\ssravg > 0$ in the half-good case.

In \figref{power_plotz_one_b}, we plot the power of the MHT and chi-bar-square procedures,
with and without Hansen's log-log corrections.
The correction has essentially no effect in the all-equal and half-good cases,
but increases power considerably for the chi-bar-square test in the two-good and one-good
alternatives, and increases power modestly for the MHT test under those same alternatives.
Thus we mostly consider using the log-log correction when possible.
We note that this correction may result in slightly higher than nominal type I rate,
as evidenced later in \figref{rho_plotz_one}.

\begin{knitrout}\small
\definecolor{shadecolor}{rgb}{0.969, 0.969, 0.969}\color{fgcolor}\begin{figure}[h]
\includegraphics[width=0.975\textwidth,height=0.691\textwidth]{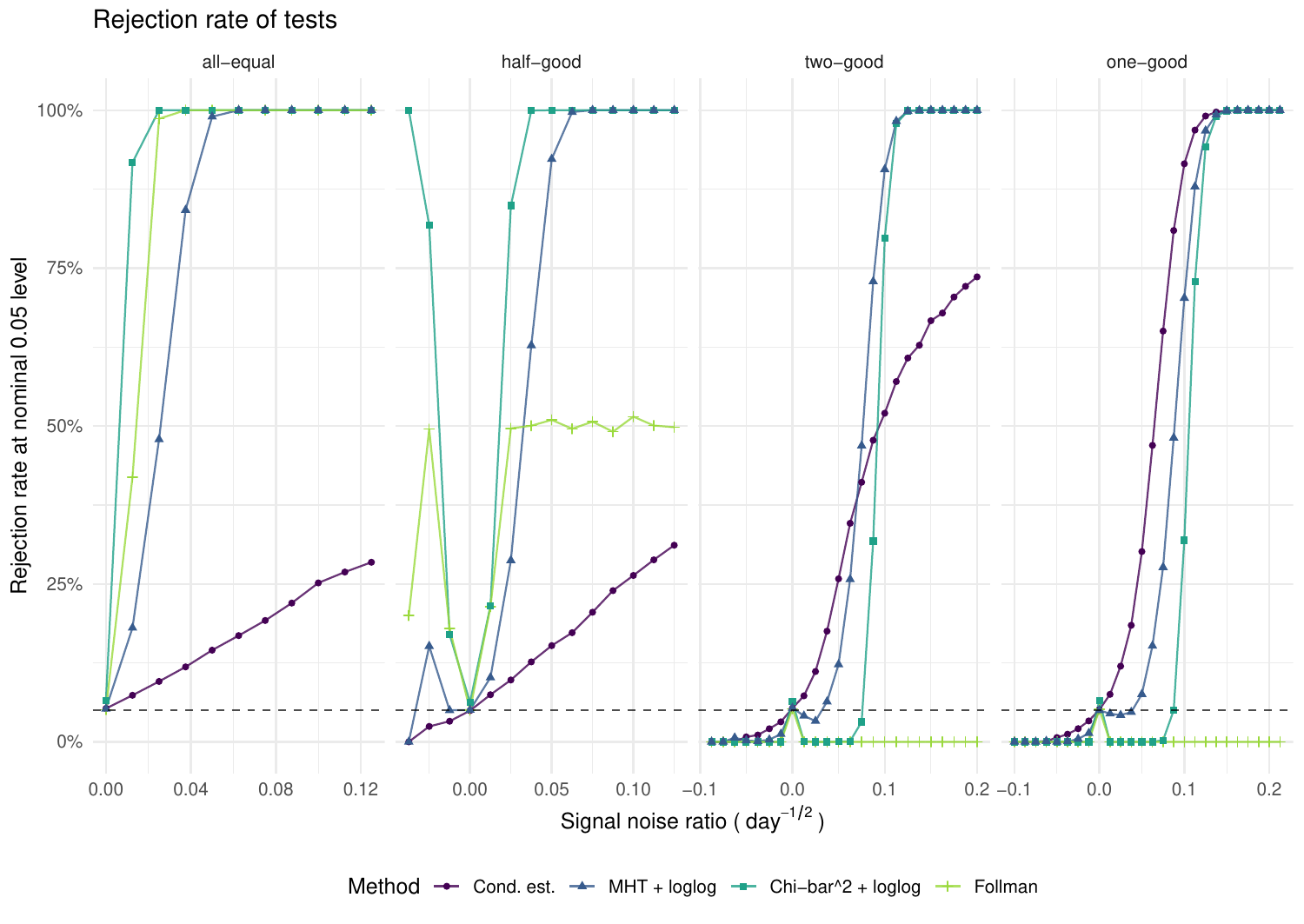} \caption[The empirical power of the conditional estimation test, Follman's test, and the chi-bar-square and MHT tests with Hansen's $\log\log$ correction are shown versus the \txtSNR of the asset with maximum \txtSR under different arrangements of the vector \pvsnr]{The empirical power of the conditional estimation test, Follman's test, and the chi-bar-square and MHT tests with Hansen's $\log\log$ correction are shown versus the \txtSNR of the asset with maximum \txtSR under different arrangements of the vector \pvsnr. For the one- and two-good cases, Follman's test has essentially zero power under the alternative. }\label{fig:power_plotz_one_a}
\end{figure}

\end{knitrout}
\begin{knitrout}\small
\definecolor{shadecolor}{rgb}{0.969, 0.969, 0.969}\color{fgcolor}\begin{figure}[h]
\includegraphics[width=0.975\textwidth,height=0.691\textwidth]{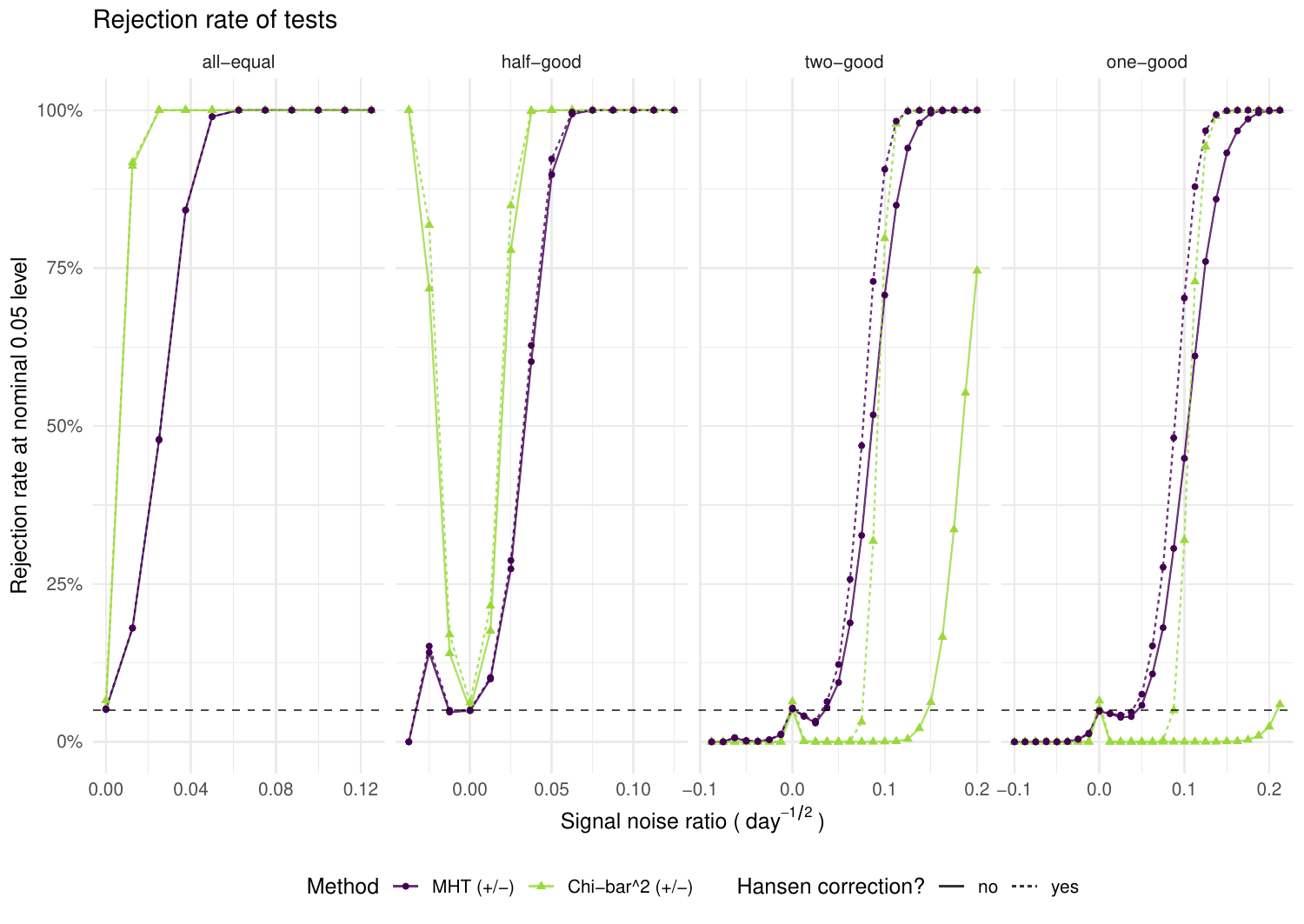} \caption[The empirical power of the MHT corrected test, the chi-bar-square test, with and without Hansen's $\log\log$ correction, are shown versus the \txtSNR of the asset with maximum \txtSR under different arrangements of the vector \pvsnr]{The empirical power of the MHT corrected test, the chi-bar-square test, with and without Hansen's $\log\log$ correction, are shown versus the \txtSNR of the asset with maximum \txtSR under different arrangements of the vector \pvsnr. }\label{fig:power_plotz_one_b}
\end{figure}

\end{knitrout}

The power of the conditional estimation procedure for the all-equal
and half-good cases is rather disappointing. 
For the case where all assets have a \txtSNR of
$3.4\yrto{-\halff}$, 
which should be considered very very large, the test has a power of only around a half.
The test suffers from low power because we are conditioning on
``\ssr[1] is the largest \txtSR'', but we should actually
like to condition on ``the asset with the largest \txtSR, whichever one that is.''
That is, there is no recognition here that if another asset 
had exhibited higher \txtSR, we would have selected that one instead.
It is not clear if that distinction can be meaningfully integrated
into this testing procedure.

However, we caution that the apparent low power of the conditional inference
test in comparison to tests based on MHT is that the latter are rejecting
a different null, and can actually have much higher type I rate under
the conditional null.
Among these tested alternatives, this effects is most visible in the
half-good facet, where
the MHT correction
and one-sided tests have greater
than $0.05$ rejection rate for \emph{negative} \txtSNR. 
Some clarification is required here.
We have performed
$\ensuremath{10^{4}}$ simulations for each setting of the `good'
\txtSNR; in some number of them for the half-good
case, an asset with negative \txtSNR exhibits the maximum
\txtSR. 
We are plotting the rejection rate for the test in this case.
But note that the null hypothesis that MHT and the one-sided test are testing
\emph{is} violated in this case, because half the assets
have positive \txtSNR, and the alternative procedures test the null that all
assets have zero or lower \txtSNR.
Tests based on the MHT do not maintain the nominal type I rate under the conditional null.
While the probability of selecting a `bad' asset instead of a `good' one is going to be very
low, especially when there is a large gap between the \txtSNRs of the good and the bad assets,
it is troubling that when this occurs, tests based on the MHT can fail with high probability.


On the other hand, while the conditional estimation procedure exhibits
lower power than the other tests (except in the one-good case),
it appears to have monotonic rejection probability with respect to the
\txtSNR of the \emph{selected} asset in all configurations tested.
That is, in the half-good case, it has low rejection probability in the odd
simulations where a `bad' asset is selected because it is actually designed
for testing the conditional null.
This pattern is repeated in the following set of simulations under the alternative.


It is unlikely that the four distributions of $\pvsnr$ considered above will be encountered in real world applications;
rather we anticipate that the elements of \pvsnr will have a distribution which is bell-shaped or even flat.
We simulate these cases, estimating the power of the various tests as a function of the \txtSNR of the selected asset.

We set $\nstrat=105$, and $\ssiz=1008$.
The elements of \pvsnr are set to fixed values shown in \figref{power_plotz_pvsnr_distribution}.
Namely for the uniform case, the \pvsnr takes uniform values from $-0.07\dayto{-\halff}$ to $0.07\dayto{-\halff}$,
or in annualized units $\ccinterval{-1.1}{1.1}\dayto{-\halff}$.
Under the bell-shaped simulations, the \pvsnr takes the same range but with more mass around zero.

\begin{knitrout}\small
\definecolor{shadecolor}{rgb}{0.969, 0.969, 0.969}\color{fgcolor}\begin{figure}[h]
\includegraphics[width=0.975\textwidth,height=0.691\textwidth]{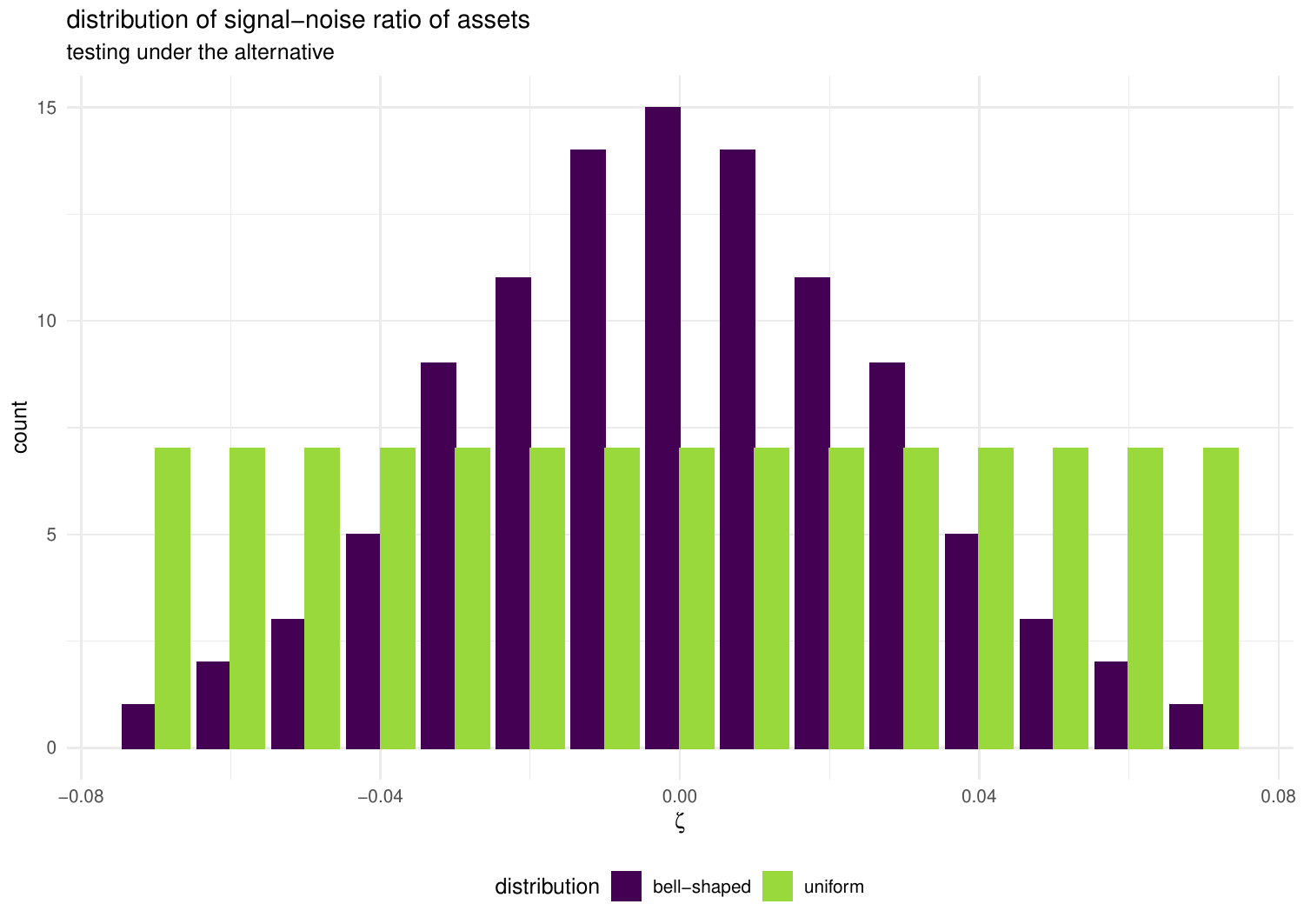} \caption[The counts of the values of \pvsnr are shown for the uniform and bell-shaped simulations under the alternative]{The counts of the values of \pvsnr are shown for the uniform and bell-shaped simulations under the alternative. The \txtSNR is given in daily units, $\dayto{-\halff}$.}\label{fig:power_plotz_pvsnr_distribution}
\end{figure}

\end{knitrout}
In \figref{power_plotz_two} we plot the empirical rejection rates for the various tests 
under these two distributions of \pvsnr.
The rejection rate is plotted against the \txtSNR of the selected strategy, \psnr[1].
We note that in some unlucky cases $\psnr[1]$ is small or even negative even though the maximum element of \pvsnr is
significantly positive.
As seen above, the conditional procedure maintains the near nominal type I rate of 0.05 when $\psnr[1] = 0$,
with modest power when $\psnr[1] > 0$.
The other tests exhibit much higher rejection rates, even when $\psnr[1] \le 0$.
This is not unexpected, as the null \emph{is} violated in this case and rejecting it is the right decision
for tests based on the MHT null.
We reiterate that tests based on the MHT can, in some minority of cases, correctly reject the null, but the 
selected strategy has unacceptable \txtSNR.
In \figref{power_plotz_two_sel_prob} we plot the empirical probability under our simulations that the
asset with maximum \txtSR has a given \txtSNR.
For both the uniform and bell-shaped \pvsnr, the probability that $\psnr[1] \le 0$ appears to be less than 10\%,
with smaller chance under the uniform distribution.
Thus the likelihood of making this kind of conditional type I error when using MHT tests is perhaps low,
it is not zero.
However, if a conditional type I error rate must be maintained, the MHT tests should be considered unacceptable.

\begin{knitrout}\small
\definecolor{shadecolor}{rgb}{0.969, 0.969, 0.969}\color{fgcolor}\begin{figure}[h]
\includegraphics[width=0.975\textwidth,height=0.691\textwidth]{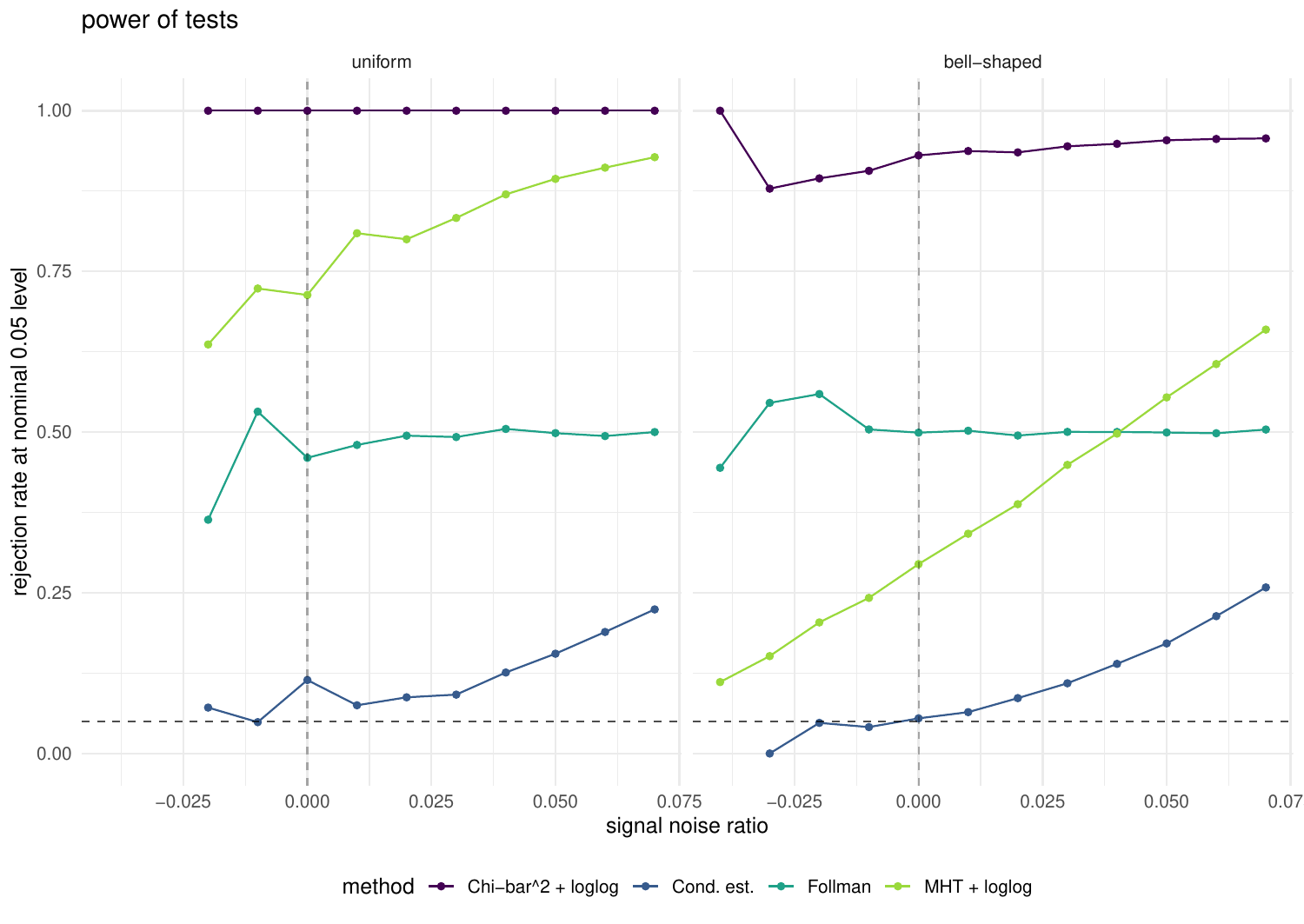} \caption[The empirical power of the conditional estimation, Follman's test, Hansen's SPA and corrected chi-bar-square procedures are shown versus the \txtSNR of the asset with maximum \txtSR for the uniform and bell-shaped arrangements of the vector \pvsnr]{The empirical power of the conditional estimation, Follman's test, Hansen's SPA and corrected chi-bar-square procedures are shown versus the \txtSNR of the asset with maximum \txtSR for the uniform and bell-shaped arrangements of the vector \pvsnr. We perform 100,000 simulations with $\ssiz=1008$, $\nstrat=105$. Cases with low incidence were removed as the simulated rejection rates are not representative. }\label{fig:power_plotz_two}
\end{figure}

\end{knitrout}
\begin{knitrout}\small
\definecolor{shadecolor}{rgb}{0.969, 0.969, 0.969}\color{fgcolor}\begin{figure}[h]
\includegraphics[width=0.975\textwidth,height=0.691\textwidth]{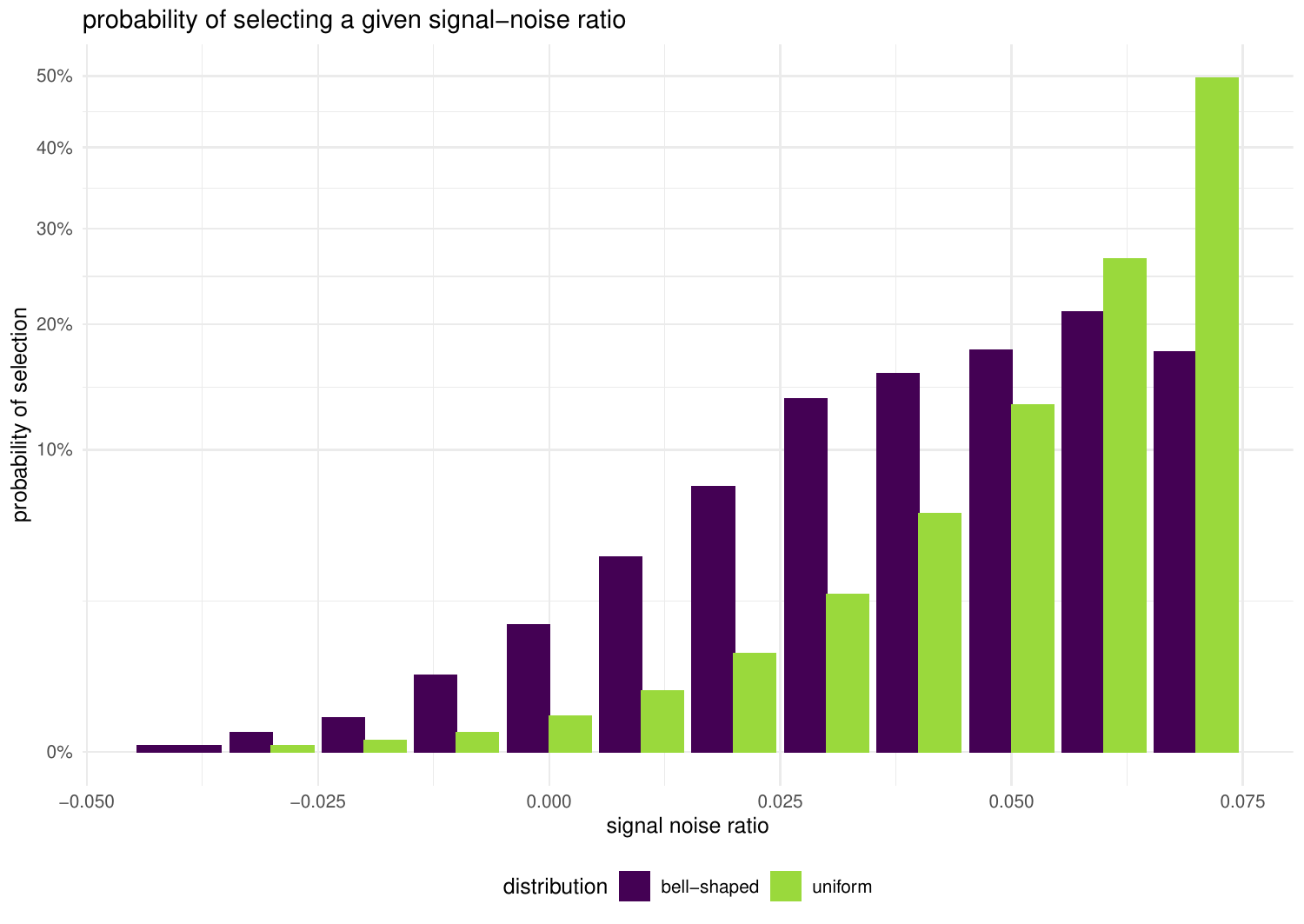} \caption[The probability that $\psnrone$, the \txtSNR of the asset selected for having the maximum \txtSR, takes a given value in our simulations is shown]{The probability that $\psnrone$, the \txtSNR of the asset selected for having the maximum \txtSR, takes a given value in our simulations is shown. The $y$ axis is in square root space to show detail.}\label{fig:power_plotz_two_sel_prob}
\end{figure}

\end{knitrout}

We doubt that the simple experiments performed here
have revealed all the relevant differences between the various tests or when
one dominates the others.

\clearpage

\subsection{Simulations of False Discovery Rate}
\label{subsec:fdr_sims}

Recalling the discussion in \secref{alternative_approaches} we now perform simulations similar to the above
but focus on the positive False Discovery Rate (pFDR). \cite{10.1111/1467-9868.00346}
We define the pFDR as the conditional probability of making a false positive
conditional on the statistical test rejecting the null.
Recalling \figref{quant_flowchart_II}, this is the quantity
\begin{equation}
pFDR = \frac{\gamma_1 + \gamma_3}{\gamma_1 + \gamma_3 + \gamma_5}.
\label{eqn:pFDR}
\end{equation}

Similar to the experiments above we consider a few different configurations for $\pvsnr$, though by no means do we consider an exhaustive collection.
In each configuration we will perform a number of simulations of the different tests considered in this paper.
We parametrize the configurations by the maximal \txtSNR, $\psnrmax$, then we will plot the results against that $\psnrmax$.
As above we consider:
\begin{compactenum}
\item ``one-good,'' where 1 asset has \txtSNR of $\psnrmax$, and the remaining $\nstrat-1$ take value of $-\psnrmax$.
\item ``half-good,'' where $\nstrat/2$ have \txtSNR of $\psnrmax$ and the remaining take value of $-\psnrmax$.
\item ``uniform,'' where the \txtSNRs are uniformly distributed across 15 different values from $-\psnrmax$ to $\psnrmax$, 
  as in \figref{power_plotz_pvsnr_distribution}.
\item ``bell-shaped,'' where the \txtSNRs are distributed across a nearly normal distribution spanning from $-\psnrmax$ to $\psnrmax$, 
  as in \figref{power_plotz_pvsnr_distribution}.
\end{compactenum}

In each configuration, except one-good, the mean \txtSNR across the different assets is zero.
We perform simulations with $\nstrat=105$, and $\ssiz=1008$.
We let $\psnrmax$ vary from near zero to around $1\yrtomhalf$.
For each setting we perform $10,000$ simulations, testing at the
$0.05$ level.
We compute the empirical pFDR of each technique.
When a technique makes no rejections we ``nan out'' the value and do not plot a marker.
This happens for the chi-bar-square test in the one-good configuration for some values of \psnrmax.

We plot the results in \figref{plot_fdr}.
In that plot we also plot the results of using the na\"{i}ve test for significance of the \txtSNR, which we denote as ``no MHT''. \cite{pav_the_book}
We also plot the pFDR resulting from always rejecting the null hypothesis, which we plot as ``Reject All.''
This line shows the background rate of bad strategies, which is $\beta_0 + \beta_1$ in the language of \figref{quant_flowchart_II}.
The Reject All line is essentially the worst pFDR that a test could display without doing something perverse.

\begin{knitrout}\small
\definecolor{shadecolor}{rgb}{0.969, 0.969, 0.969}\color{fgcolor}\begin{figure}[h]
\includegraphics[width=0.975\textwidth,height=0.682\textwidth]{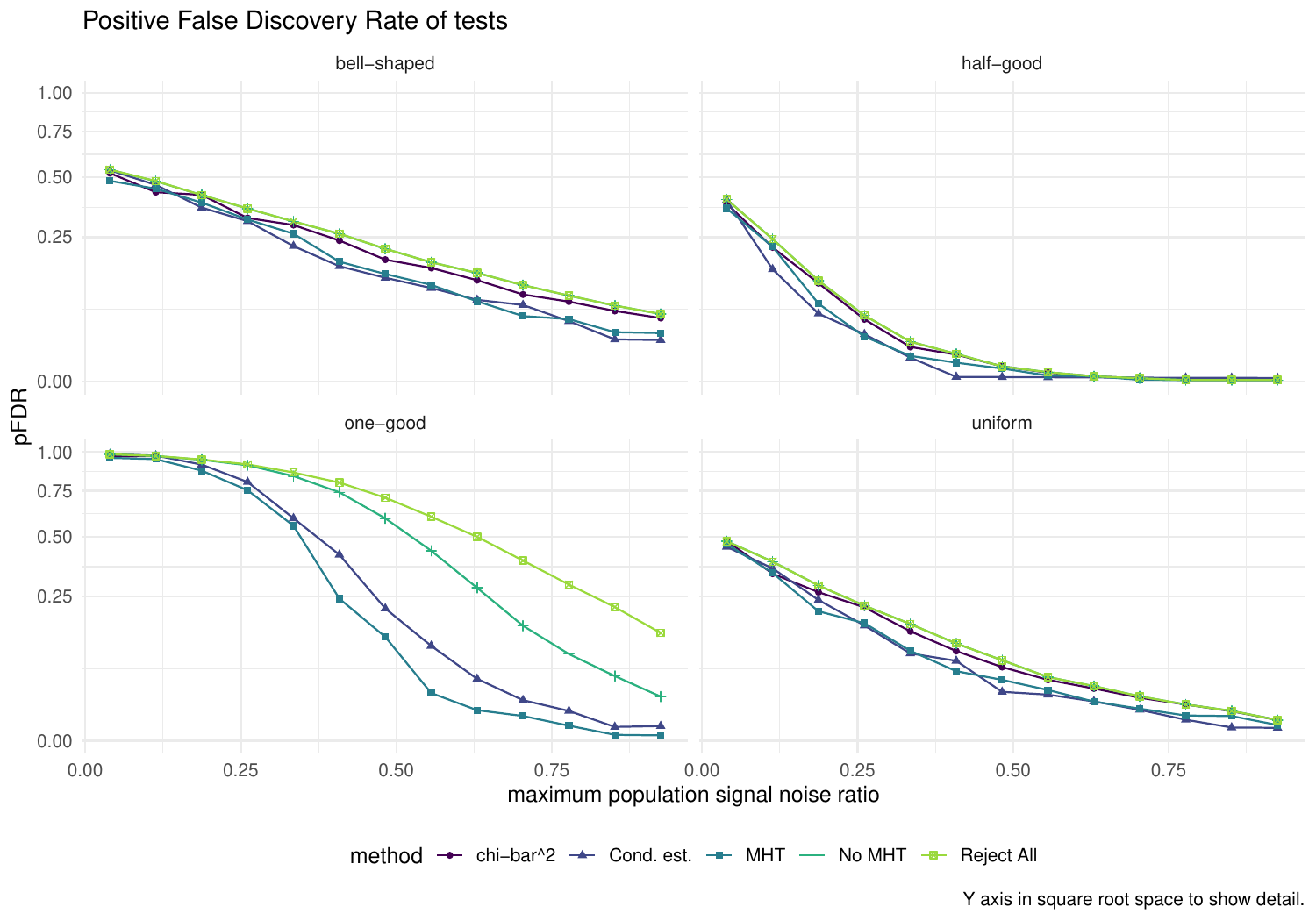} \caption[The empirical positive False Discovery Rate (pFDR) of the various tests are shown versus the maximum population \txtSNR, \psnrmax, for various configurations of the true \txtSNR]{The empirical positive False Discovery Rate (pFDR) of the various tests are shown versus the maximum population \txtSNR, \psnrmax, for various configurations of the true \txtSNR. We perform 10,000 simulations with $\ssiz=1008$, $\nstrat=105$. The chi-bar-square test has no rejections for most of the one-good case, and is not plotted there. }\label{fig:plot_fdr}
\end{figure}

\end{knitrout}

We see that MHT and the conditional procedure generally have the lowest pFDR,
generally lower than the chi-bar-square procedure, lower than the na\"{i}ve test, and lower than the Reject All benchmark.
The MHT performs best in the one-good configuration.
We perform some additional simulations under the uniform and bell-shaped configurations,
since these seem more realistic thatn the one-good and half-good.
We perform 100,000 simulations for these configurations,
plotting the results in \figref{plot_fdr_II}.
There we see that the conditional inference procedure has slightly lower pFDR than the MHT procedure
when \psnrmax is sufficiently large.
It is hard to imagine that this result scales to all sample sizes and configurations,
but we are generally confident that the two procedures deliver similar pFDR.

\begin{knitrout}\small
\definecolor{shadecolor}{rgb}{0.969, 0.969, 0.969}\color{fgcolor}\begin{figure}[h]
\includegraphics[width=0.975\textwidth,height=0.682\textwidth]{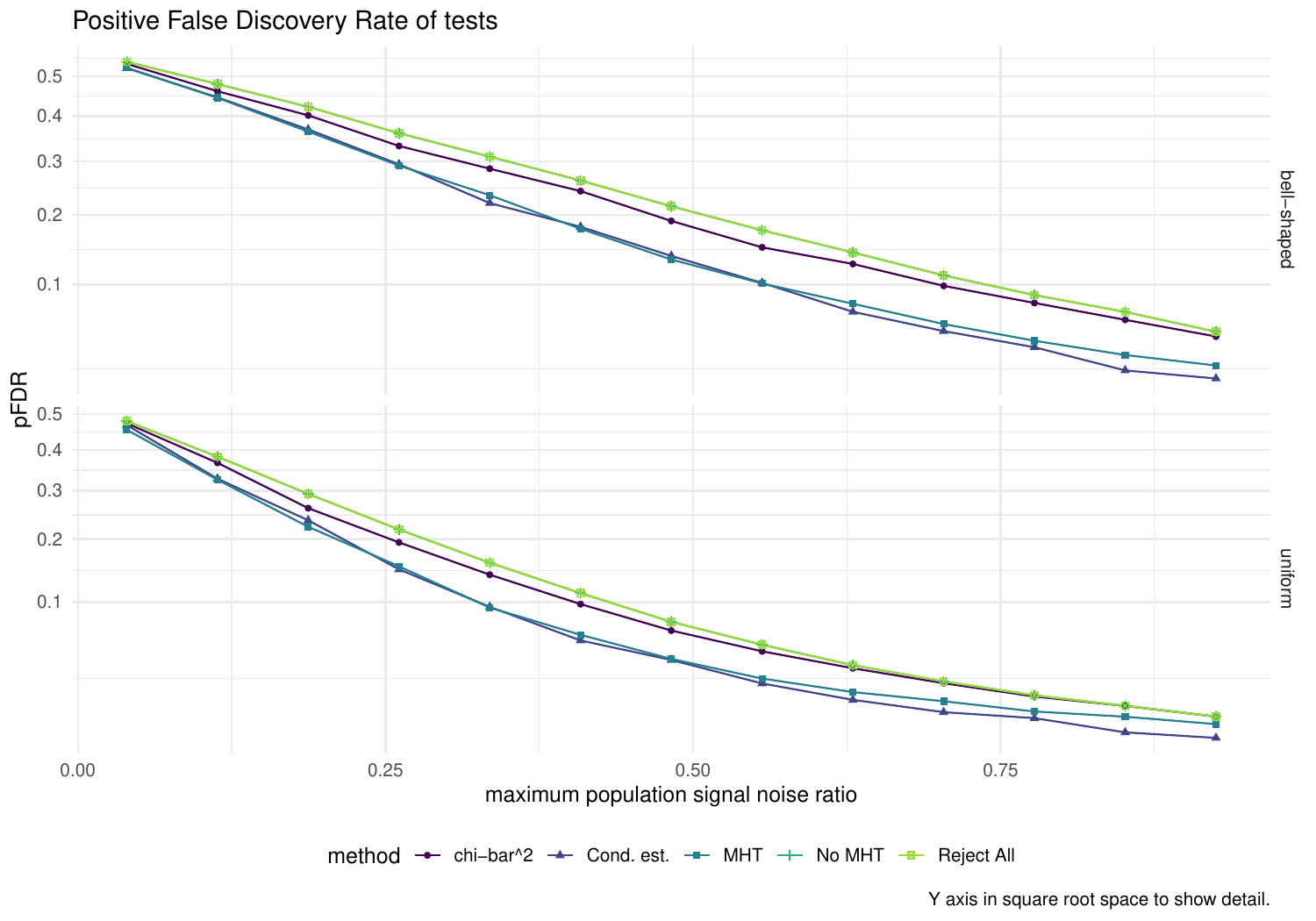} \caption[The empirical positive False Discovery Rate (pFDR) of the various tests are shown versus the maximum population \txtSNR for various arrangements of the true \txtSNR]{The empirical positive False Discovery Rate (pFDR) of the various tests are shown versus the maximum population \txtSNR for various arrangements of the true \txtSNR. We perform 100,000 simulations with $\ssiz=1008$, $\nstrat=105$. }\label{fig:plot_fdr_II}
\end{figure}

\end{knitrout}

\subsubsection{Minimizing False Discovery Rate}
\label{subsec:minimizing_fdr}

As a thought experiment, let us consider how we might minimize the pFDR for a toy problem.
Let $\pi$ be the probability that the selected strategy satisfies the null hypothesis,
which we define to mean having \txtSNR of zero.
We have $\pi = \beta_0 + \beta_1$ in the language of \figref{quant_flowchart_II}.
Suppose that the type I rate under the null is $\alpha$.
Under the alternative, which occurs with probability $1 - \pi$, the strategy is ``good''
and has a \txtSNR of \psnrmax.
Suppose the observed \txtSR is the \txtSNR plus a normal noise with unit variance.
So our test consists of rejecting the null precisely when the \txtSR is greater than
$\pinorm[1-\alpha]$, where \pnorm[x] is the cumulative density of the standard normal distribution.
Under the alternative this occurs with probability $1 - \pnorm[\pinorm[1-\alpha] - \psnrmax]$.

Then the pFDR is 
\begin{align*}
pFDR 
  &= \frac{\pi \alpha}{\pi \alpha + \wrapParens{1 - \pi}\wrapParens{1 - \pnorm[\pinorm[1-\alpha] - \psnrmax]}},\\
  &= \frac{1}{1 + \frac{\wrapParens{1 - \pi}\wrapParens{\pnorm[\psnrmax - \pinorm[1-\alpha]]}}{\pi\alpha}},\\
  &= \frac{1}{1 + \frac{\wrapParens{1 - \pi}\wrapParens{\pnorm[\psnrmax + \pinorm[\alpha]]}}{\pi\pnorm[\pinorm[\alpha]]}}.\\
\end{align*}
Without asking the analyst to improve $\pi$ by testing only good strategies, to minimize the pFDR we should 
maximize the quantity
$$
\frac{\pnorm[\psnrmax + \pinorm[\alpha]]}{\pnorm[\pinorm[\alpha]]}.
$$
Linearizing this with Taylor's theorem we want to approximately maximize the quantity
$$
\frac{\pnorm[\psnrmax + \pinorm[\alpha]]}{\pnorm[\pinorm[\alpha]]} 
\approx 
1 + \psnrmax \frac{\dnorm[\pinorm[\alpha]]}{\pnorm[\pinorm[\alpha]]}.
$$
Note that the quantity $\dnorm[x]/\pnorm[-x]$ is called the \emph{inverse Mills ratio}.
It is well known that the inverse Mills ratio is unbounded and increasing as $x \to \infty$.
The implication for our problem is that pFDR is minimized as we drive the type I rate to zero.

While we cannot literally set $\alpha=0$, since we would never reject the null hypothesis,
it is easy to reconsider the experiments above using different type I rates.
That is exactly what we do: we take the MHT and conditional test results from the simulations
above, wherein we computed their p-values, and compare them against different target type I rates,
estimating the pFDR in each case.
We plot these in \figref{plot_fdr_III}.
To show detail in this plot, we have divided the pFDR by the ``worst case'' pFDR one gets from the
Reject All test, which is to say the background rate of bad strategies.

We indeed see in this plot that setting the type I rate to 0.005 yields the lowest pFDR,
and that the pFDR values are generally monotonic in the targeted type I rate.
We see that at the most extreme values of $\alpha$ that the conditional procedure has slightly lower pFDR
than the MHT test, but again this is not an exhaustive study of the possible configurations.
We would recommend, however, that if one seeks to minimize pFDR the choice of $\alpha$ is likely more important than which test you apply.

\begin{knitrout}\small
\definecolor{shadecolor}{rgb}{0.969, 0.969, 0.969}\color{fgcolor}\begin{figure}[h]
\includegraphics[width=0.975\textwidth,height=0.682\textwidth]{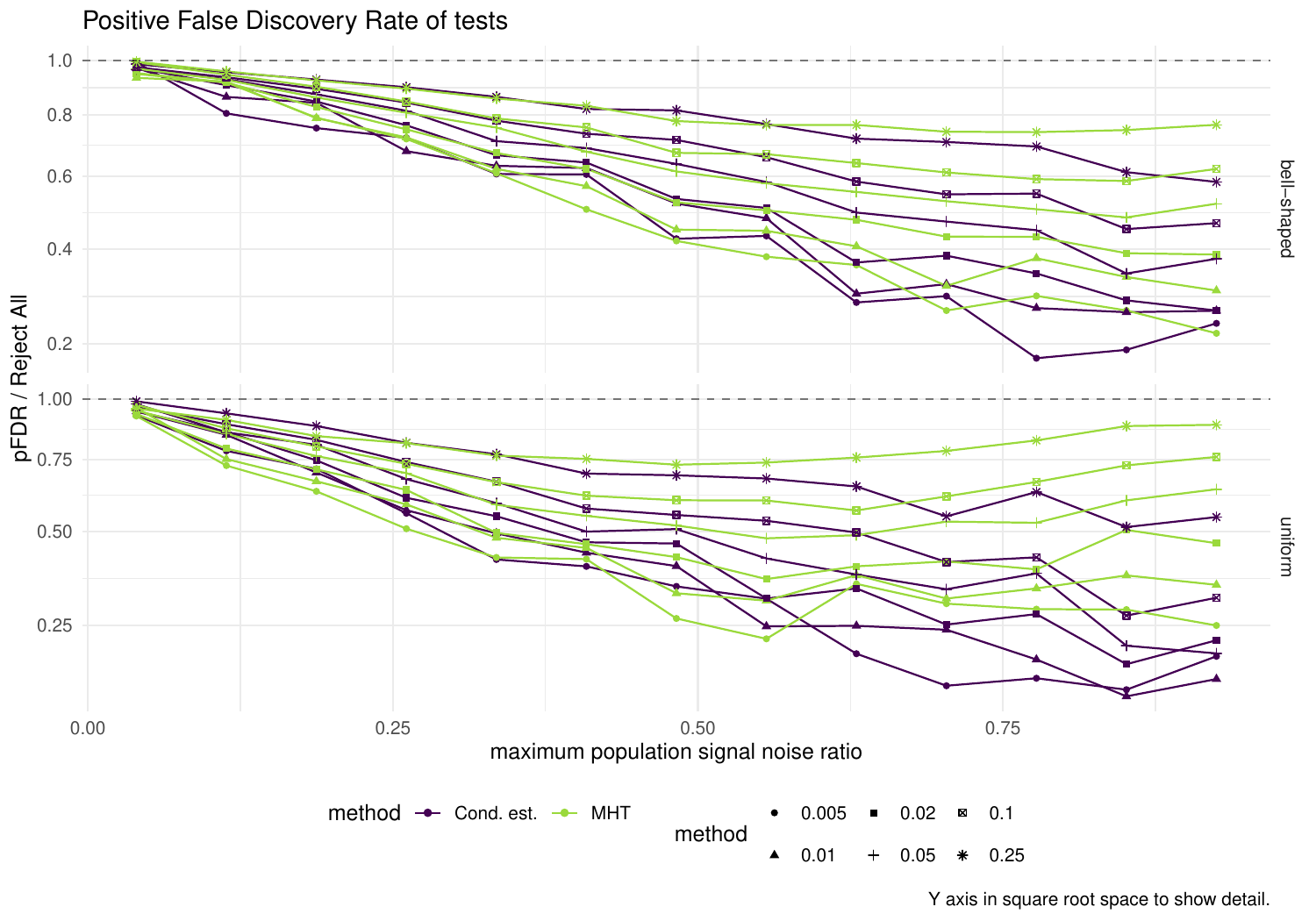} \caption[The empirical positive False Discovery Rate (pFDR) of the conditional estimation and MHT procedures, divided by the Reject All pFDR are plotted versus the maximum population \txtSNR, \psnrmax, for various configurations of the true \txtSNR]{The empirical positive False Discovery Rate (pFDR) of the conditional estimation and MHT procedures, divided by the Reject All pFDR are plotted versus the maximum population \txtSNR, \psnrmax, for various configurations of the true \txtSNR. We perform 100,000 simulations with $\ssiz=1008$, $\nstrat=105$. }\label{fig:plot_fdr_III}
\end{figure}

\end{knitrout}

\clearpage
\subsection{Simulations under the null with correlated returns}

The na{\"i}ve MHT test cannot maintain the nominal type I rate in the face of correlated assets.
To demonstrate this, we repeat the simulations above, testing the various statistical tests, but with
$\RMAT=\makerho{\rho}{\wrapParens{1-\rho}}$
and $\psnr=\vzero$.
We set $\nstrat=20$, $\ssiz=504$, 
and perform $\ensuremath{10^{4}}$ simulations to estimate the empirical rejection rate.
In \figref{rho_plotz_one}, we plot the empirical rejection rate versus $\rho$ at the nominal
$0.05$ type I level. 
While the conditional estimation procedure and one-sided tests appear to maintain the nominal rejection
rate, 
the MHT test is conservative, with near zero rejection rates for large $\rho$.
The fix for common correlation described in \subsecref{fix_bonferroni} is also tested,
yielding nominal rejection rates.

\begin{knitrout}\small
\definecolor{shadecolor}{rgb}{0.969, 0.969, 0.969}\color{fgcolor}\begin{figure}[h]
\includegraphics[width=0.975\textwidth,height=0.691\textwidth]{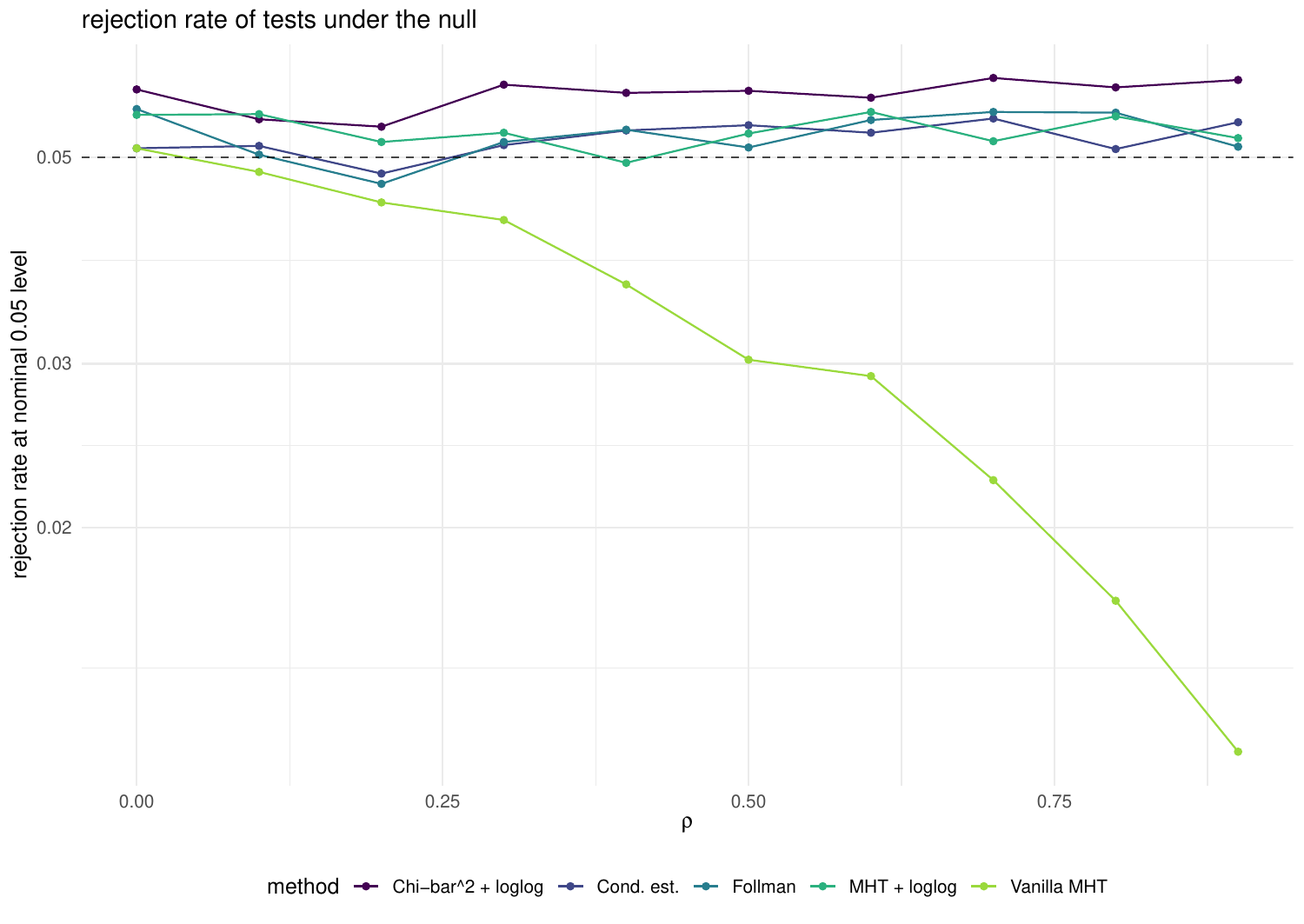} \caption[The empirical type I rate under the null hypothesis is plotted versus $\rho$ for the case where $\RMAT=\makerho{\rho}{\wrapParens{1-\rho}}$, for several testing procedures]{The empirical type I rate under the null hypothesis is plotted versus $\rho$ for the case where $\RMAT=\makerho{\rho}{\wrapParens{1-\rho}}$, for several testing procedures: the conditional estimation, the chi-bar-square test, Follman's test, the vanilla Bonferroni correction, Hansen's corrected chi-bar-square and SPA tests, and Bonferroni correction with fix for common correlation.  Tests are performed with Gaussian returns, for 20 assets over 504 days. Tests were performed at the $0.05$ level, which appears to be maintained by all procedures except the regular Bonferroni procedure, and Hansen's chi-bar-square test, which has slightly higher than nominal rejection rates for this \ssiz. Empirical rates are over 10,000 simulations. The $y$ axis is in log scale.}\label{fig:rho_plotz_one}
\end{figure}

\end{knitrout}

\clearpage

\subsection{Example: Five Industry Portfolios}


We download the monthly returns of five industry portfolios
from Ken French's data library. \cite{ind_5_def}
We consider the 1187 months of data on five industries
from Jan 1927 through Nov 2025.
We compute the \txtSR of the returns of each, and present them
in \tabref{mind_sr}. 
We have reordered the industries in decreasing \txtSR.
The industry portfolio with the highest \txtSR was
Healthcare with a \txtSR of 
around $0.193\,\moto{-\halff}$ which is approximately
$0.669\,\yrto{-\halff}$.

\begin{table}[ht]
\centering
\begin{tabular}{r|c}
  \hline
industry & Sharpe Ratio \\ 
  \hline
Healthcare & $ 0.193\,\moto{-\halff} $ \\ 
  Consumer & $ 0.192\,\moto{-\halff} $ \\ 
  Technology & $ 0.181\,\moto{-\halff} $ \\ 
  Manufacturing & $ 0.174\,\moto{-\halff} $ \\ 
  Other & $ 0.144\,\moto{-\halff} $ \\ 
   \hline
\end{tabular}
\caption{The \txtSRs of the five industry portfolios are shown.} 
\label{tab:mind_sr}
\end{table}

We are interested in computing $95\%$ upper confidence
intervals on the \txtSNR of the Healthcare portfolio. 
We are only considering this portfolio as it is the one with maximum \txtSR
in our sample.
If we had been interested in testing
Healthcare 
without our conditional selection, we would compute the confidence interval
$\left[0.145\,\moto{-\halff}, \infty\right)$ based on inverting the non-central
\tstat{}-distribution. \cite{pav_ssc,pav_the_book,SharpeR-Manual}
If instead we approximate the standard error by plugging in 
$0.193\,\moto{-\halff}$ as the \txtSNR of 
Healthcare 
into \eqnref{apx_srdist_gaussian},
we estimate the standard error of the \txtSR to be
$0.029\,\moto{-\halff}$.
Based on this we can compute the na\"{i}ve confidence interval of
the measured \txtSR plus 
$\qnorm{0.05}=-1.645$ times the standard error.
This also gives the confidence interval
$\left[0.145\,\moto{-\halff}, \infty\right)$.

Using the simple Bonferroni correction, however, since we selected
Healthcare 
only for having the maximum \txtSR, we should compute the 
confidence interval by adding
$\qnorm{0.01}=-2.326$ times the standard error.
This yields the confidence interval
$\left[0.125\,\moto{-\halff}, \infty\right)$.

The correlation of industry returns is high, however.
The pairwise sample correlations range from
0.699 to 
0.892 with a median value of
0.793.
Plugging this value in as $\rho$, 
we find the value of \psnr[0] such that the $z_1$ from \eqnref{fix_bonf_z}
is equal to 
$\qnorm{0.01}=-2.326$.
This leads to the confidence interval 
$\left[0.13\,\moto{-\halff}, \infty\right)$.

We use this estimate of $\rho$ to compute the chi-bar-square test.
We invert the test to find the $95\%$ upper confidence interval
$\left[0.146\,\moto{-\halff}, \infty\right)$.
These two intervals invert tests of the null of equality, and need not be valid
under (\ref{hyp:ineq_test}).

Finally we use the conditional estimation procedure, inverting
the hypothesis test to find the corresponding population value.
This yields the confidence interval 
$\left[-0.056\,\moto{-\halff}, \infty\right)$.
This is a much wider interval which encompasses zero.
The extra width is not surprising given the results of our simulations,
and is the price we pay for testing the selected asset conditionally on the observed.

\section{Conclusions and Future Work}

The conditional estimation procedure appears to achieve nominal
type I rates under the null, and does not seem unduly harmed by assuming
the vector \svsr is normally distributed. 
Nor does it seem to suffer greatly from using sample estimates
of the correlation matrix, \RMAT, nor from the presence of 
kurtotic returns.
The procedure can be used for other test configurations beyond the
asset with the maximum \txtSR,
and can be used to construct confidence intervals.
It appears to have low power compared to tests based on the MHT for
some test configurations.
However MHT tests can have very high conditional type I rates in those unlikely 
cases where an asset with low \txtSNR is selected,
while the conditional estimation procedure maintains the type I rate in this case.

The low power of the test gives us reason to seek improvements.
Perhaps the conditioning procedure can be adapted to recognize
that the strategist would have been testing another asset if the
\txtSR of the currently selected asset had been lower. 
However, it would seem that in so doing, one would lose the
desireable property of testing the \txtSNR of the selected
asset, instead of testing the more general hypothesis of (\ref{hyp:eq_test}). 
Perhaps its power can be increased using Hansen's $\log\log$ adjustment.

On the other hand, the conditional procedure appears to have low positive False Discovery Rates,
and is competitive with the other tests on this metric.
We suggest setting low target type I rates to minimize the pFDR.
Perhaps the procedure of \citeauthor{10.1111/1467-9868.00346} can be adapted to this problem 
to estimate the pFDR for a given type I rate.  \cite{10.1111/1467-9868.00346} 


\bibliographystyle{plainnat}
\bibliography{common}

\appendix

\section{AI Disclosure}

We used Claude Opus 5.5 to check later revisions of this work, including:

\begin{compactenum}
\item Checking consistency of the math.
\item Confirming the code matched the math.
\end{compactenum}

A number of changes followed from this:

\begin{compactenum}
\item The model identified a flaw in our claims about the correction for common correlation.
This caused us to remove or soften a number of claims which would not hold up when some assets had negative \txtSNR.
\item Similarly, the model flagged an error in our claimed use of Slepian's lemma under the correlation correction.
\item The model also identified that we had reused a constant $c$ to mean different things in different subsections, which would have
been confusing to a reader. We change the variable to clarify.
\end{compactenum}

All remaining errors are those of the human author.

\section{Establishing \eqnref{apx_srdist_elliptical}.}

Let \Mtx{V} be the variance covariance to be computed. Then from
\eqnref{delmethod}, 
\begin{align*}
  \Mtx{V} &= \qoform{\pvvar}{\wrapParens{\dbyd{\pvsnr}{\vcat{\pvmu}{\pvmom[2]}}}},\\
   &= 
   \Mdiag{\frac{1}{\pvsigma^2}}
\onebytwo{\Mdiag{\pvsigma + \pvmu\pvsnr}}{
  {\Mdiag{\frac{- \pvsnr}{2}}}} 
  \pvvar
\twobyone{\Mdiag{\pvsigma + \pvmu\pvsnr}}{
  {\Mdiag{\frac{- \pvsnr}{2}}}} 
   \Mdiag{\frac{1}{\pvsigma^2}}.
\end{align*}

Plugging in \pvvar from \eqnref{elliptical_variances}, we have
\begin{align*}
  \Mtx{V} &= 
   \Mdiag{\frac{1}{\pvsigma^2}}
   \left\{
\Mdiag{\pvsigma + \pvmu\pvsnr}
\pvsig
\Mdiag{\pvsigma + \pvmu\pvsnr}
+ 
\Mdiag{\pvsigma + \pvmu\pvsnr}
2\pvsig\Mdiag{\pvmu}
\Mdiag{\frac{- \pvsnr}{2}}
   \right.\\
   &\phantom{=\Mdiag{\frac{1}{\pvsigma^2}}}\,
   \left.
+\Mdiag{\frac{- \pvsnr}{2}}
2\Mdiag{\pvmu}\pvsig
\Mdiag{\pvsigma + \pvmu\pvsnr}
+
\Mdiag{\frac{- \pvsnr}{2}}
2 \pvsig \hadm \pvsig
\Mdiag{\frac{- \pvsnr}{2}}\right.\\
   &\phantom{=\Mdiag{\frac{1}{\pvsigma^2}}}\,
   \left.
+
\Mdiag{\frac{- \pvsnr}{2}}
	4 \Mdiag{\pvmu}\pvsig\Mdiag{\pvmu}
\Mdiag{\frac{- \pvsnr}{2}}
   \right\}
   \Mdiag{\frac{1}{\pvsigma^2}},\\
  &= 
   \Mdiag{\frac{1}{\pvsigma^2}}
   \left\{
 \Mdiag{\pvsigma}
\pvsig
  \Mdiag{\pvsigma}
-
\Mdiag{\pvmu\pvsnr}
\pvsig
\Mdiag{\pvmu\pvsnr}\right.\\
   &\phantom{=\Mdiag{\frac{1}{\pvsigma^2}}}\,
   \left.
+
\Mdiag{\frac{- \pvsnr}{2}}
2 \pvsig \hadm \pvsig
\Mdiag{\frac{- \pvsnr}{2}}\right.\\
   &\phantom{=\Mdiag{\frac{1}{\pvsigma^2}}}\,
   \left.
+
\Mdiag{\pvmu\pvsnr}
\pvsig
\Mdiag{\pvmu\pvsnr}
   \right\}
   \Mdiag{\frac{1}{\pvsigma^2}},\\
&=
   \Mdiag{\frac{1}{\pvsigma^2}}
   \wrapBraces{
 \Mdiag{\pvsigma}
\pvsig
  \Mdiag{\pvsigma}
  + \half
\Mdiag{\pvsnr}
\pvsig \hadm \pvsig
\Mdiag{\pvsnr}}
   \Mdiag{\frac{1}{\pvsigma^2}},\\
&=
\RMAT 
+ \half
\Mdiag{\pvsnr}
   \Mdiag{\frac{1}{\pvsigma^2}}
\pvsig \hadm \pvsig
   \Mdiag{\frac{1}{\pvsigma^2}}
\Mdiag{\pvsnr},\\
&=
\RMAT 
+ \half
\Mdiag{\pvsnr}
\RMAT \hadm \RMAT
   \Mdiag{\frac{1}{\pvsigma^2}}.
\end{align*}

Proving \eqnref{apx_srdist_elliptical} is similar, and is left as an exercise for the reader.

\end{document}